\documentclass[a4paper,twoside]{ociamthesis}
\correctionstrue

\title{Chemical Reaction Dynamics under Vibrational Strong Coupling}
\author{Andrew C. Hunt}
\college{Pembroke College}
\degree{Honour School of Chemistry}
\degreedate{Part II 2023}

\usepackage{import}
\usepackage{amsmath}
\usepackage{physics}
\usepackage{graphicx}
\usepackage{mathrsfs}
\usepackage{siunitx}
\usepackage[section]{placeins}
\usepackage{upgreek}
\usepackage{dsfont}
\usepackage{caption}
\usepackage{subcaption}

\newcommand\dubarrow{%
        \mathrel{\vcenter{\mathsurround0pt
                \ialign{##\crcr
        \noalign{\nointerlineskip}$\xrightarrow{\hspace{25pt}}$\crcr
        \noalign{\nointerlineskip}$\xleftarrow{\hspace{25pt}}$\crcr
                }%
        }}%
}
    \newcommand{\vc}{\boldsymbol}
    \newcommand{\mt}{\boldsymbol}
    \newcommand{\1}{\hat{\mathds{1}}}
    \newcommand{\intinf}{\int_{\!-\infty}^{\infty}\!\!}
\begin{document}

%%%%% CHOOSE YOUR LINE SPACING HERE
% This is the official option.  Use it for your submission copy and library copy:
\setlength{\textbaselineskip}{26pt}
% This is closer spacing (about 1.5-spaced) that you might prefer for your personal copies:
%\setlength{\textbaselineskip}{18pt plus2pt minus1pt}

% You can set the spacing here for the roman-numbered pages (acknowledgements, table of contents, etc.)
\setlength{\frontmatterbaselineskip}{17pt plus1pt minus1pt}

% Leave this line alone; it gets things started for the real document.
\setlength{\baselineskip}{\textbaselineskip}

%%%%% CHOOSE YOUR SECTION NUMBERING DEPTH HERE
% You have two choices.  First, how far down are sections numbered?  (Below that, they're named but
% don't get numbers.)  Second, what level of section appears in the table of contents?  These don't have
% to match: you can have numbered sections that don't show up in the ToC, or unnumbered sections that
% do.  Throughout, 0 = chapter; 1 = section; 2 = subsection; 3 = subsubsection, 4 = paragraph...

% The level that gets a number:
\setcounter{secnumdepth}{2}
% The level that shows up in the ToC:
\setcounter{tocdepth}{2}

% JEM: Pages are roman numbered from here, though page numbers are invisible until ToC.  This is in
% keeping with most typesetting conventions.
\begin{romanpages}

% Title page is created here
\maketitle

%%%%% DEDICATION -- If you'd like one, un-comment the following.
%\begin{dedication}
%This thesis is dedicated to\\
%someone\\
%for some special reason\\
%\end{dedication}

%%%%% ACKNOWLEDGEMENTS -- Nothing to do here except comment out if you don't want it.
\begin{acknowledgements}
 	% There was never a dull moment working with the Manolopoulos group. 
Deep in the bowels of the PTCL theory wing there lies an oasis of stimulating conversation and hilarious moments, all thanks to some of the most unique and inspiring people I have ever had the pleasure of working alongside. 
% None of these however would be present without the wonderfull people that have surrounded me in this past year.
\\
\indent 
Firstly, I would like to thank my supervisor David Manolopoulos for his guidance throughout this year. Watching you formulate entire arguments in your head - with clarity only paralleled by the signature handwriting employed in their execution - has truly been incredible. The meetings between you, me and Johan, to which you gave the apt name ``\textit{bullshit hour}'' will forever be amongst my favorite memories from my time at Oxford - thank you. 
% It was so generous of you and Johan to give your time, and provided me with such valuable experiences - thank you.
\\
\indent
If one were to be given the tools to design the perfect mentor they would struggle to recreate someone anywhere near as kind or knowledgeable as Johan Runeson. You have broadened my understanding in physics, rectified many of my mathematical misconceptions and most importantly shown me unwavering patience, for which I can only express my utmost gratitude.
\\
\indent 
Annina and Seth - I am amazed you have managed to put up with me these past nine months. Thanks for the interesting conversations and teaching me all I know now about the dark arts of FORTRAN77 debugging. Aadi, Andrew, Daniel, Chris and Alex: if I had more space, you too would have paragraphs... Thanks you all for the jokes, thoughts and memories upon which I will forever reminisce.
\\
\indent 
Finally, to my family and friends of which did not reside with me in the theory wing, thank you for letting me bore you with the latest details of my project, and the support you have given me throughout the ups and downs of this truly epic year.
\end{acknowledgements}

%%%%% SUMMARY -- Nothing to do here except comment out if you don't want it.
\begin{abstract}
	In this thesis, we use classical, semi-classical and quantum-mechanical methods to simulate chemical reaction dynamics inside of an optical cavity. Within such a cavity, by selectively coupling vibrational modes of the reactants to the vacuum state of light, recent experiments have observed significant changes in reaction rates and equilibrium constants - all without any external input of energy.
\\
\indent 
We investigate the dynamics of both a single reaction and an ensemble of $N$ identical reactions coupled to the cavity. In our single reactant studies, we find significant modification to the rate of reaction and to its quantum-mechanical equilibrium constant.
\\
\indent 
All of the effects observed in our single molecule studies are however found to diminish as the number of reactants is increased. For any experimentally relevant number of molecules, the cavity effects on the reaction rate and the equilibrium constant are therefore shown to be negligible within all theories considered in this thesis. 
\\
\indent 
This thesis therefore does not offer any explanation for the experimental observations. It does however highlight issues with all current theoretical work on this topic, and provides suggestions - in light of the results presented here and in recent literature - as to what might be required to explain these effects.
\end{abstract}

%%%%% MINI TABLES
% This lays the groundwork for per-chapter, mini tables of contents.  Comment the following line
% (and remove \minitoc from the chapter files) if you don't want this.  Un-comment either of the
% next two lines if you want a per-chapter list of figures or tables.
\dominitoc % include a mini table of contents
%\dominilof  % include a mini list of figures
%\dominilot  % include a mini list of tables

% This aligns the bottom of the text of each page.  It generally makes things look better.
\flushbottom

% This is where the whole-document ToC appears:
\tableofcontents

\listoffigures
	\mtcaddchapter
% \mtcaddchapter is needed when adding a non-chapter (but chapter-like) entity to avoid confusing minitoc

% Uncomment to generate a list of tables:
%\listoftables
%	\mtcaddchapter

%%%%% LIST OF ABBREVIATIONS
% This example includes a list of abbreviations.  Look at text/abbreviations.tex to see how that file is
% formatted.  The template can handle any kind of list though, so this might be a good place for a
% glossary, etc.
% First parameter can be changed eg to "Glossary" or something.
% Second parameter is the max length of bold terms.
\begin{mclistof}{List of Abbreviations}{3.2cm}

% \item[Fig.] Figure
% \item[Ch.] Chapter
% \item[Ref.] Reference
\item[VSC] Vibrational strong coupling
\item[PTA]  1-phenyl-2-trimethylsilylacetylene
\item[IR] Infrared
\item[HTST] Harmonic transition state theory
\item[TST] Transition state theory
\item[RPMD] Ring polymer molecular dynamics
\item[MD] Molecular dynamics
\item[ZPE] Zero-point energy
\item[QUAPI] Quasi-adibatic propagator path integral
\item[HEOM] Hierachical equations of motion
\item[GST] Ground-state tunneling
\item[GLE] Generalised Langevin equations

\end{mclistof}

% The Roman pages, like the Roman Empire, must come to its inevitable close.
\end{romanpages}

%%%%% CHAPTERS
% Add or remove any chapters you'd like here, by file name (excluding '.tex'):
% \begin{savequote}[8cm]
% `Polariton chemistry is real, and should be taken seriously - just like quantum computing!'
%   \qauthor{--- David E. Manolopolous }
% \end{savequote}
\chapter{\label{1:}Introduction} 
\indent 
Synthetic chemists have long sought after ways to improve the selectivity and yield of their chemical transformations. Recently, a new method has emerged promising to achieve both of these features by selectively coupling vibrational modes of reactive systems to the vacuum state of light. This is known as chemistry under vibrational strong coupling (VSC).
\begin{figure}
\centering
\begin{subfigure}[c]{0.485\textwidth}
  \centering
\scalebox{1.}{\includegraphics[width=\textwidth]{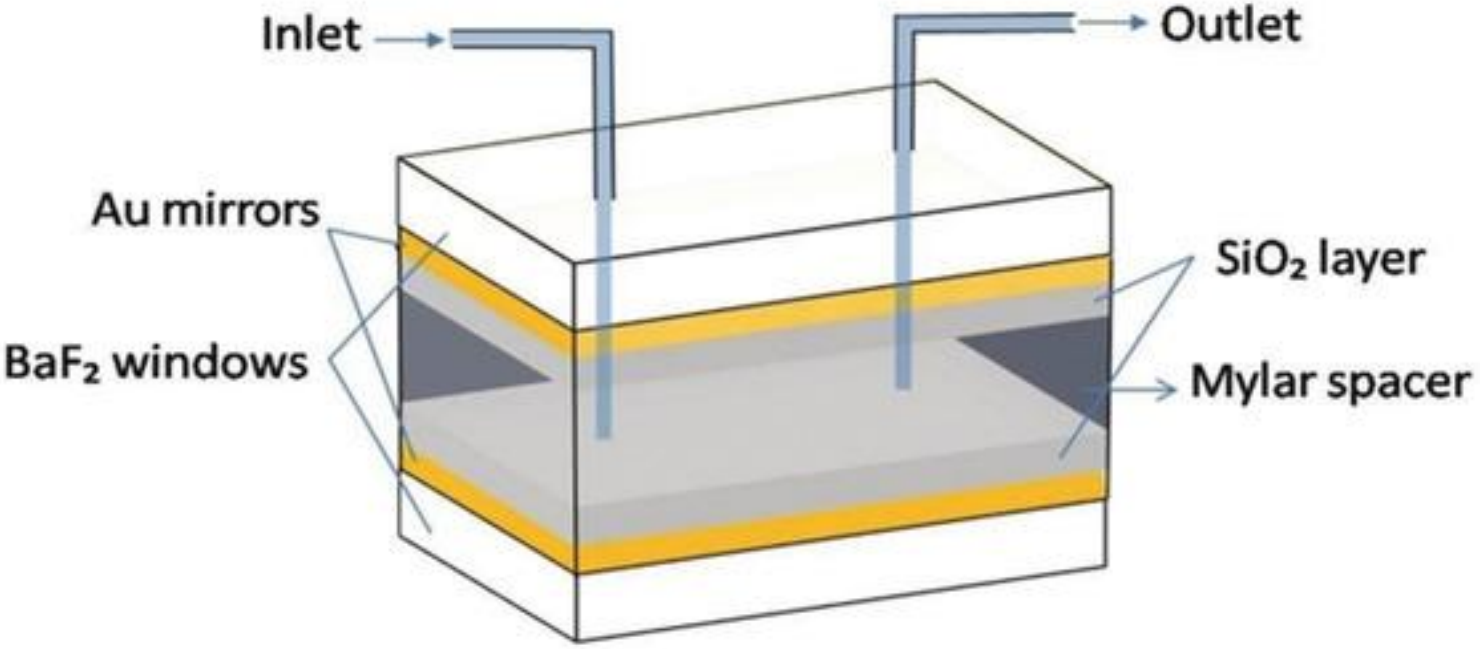}}
\end{subfigure}
\hfill
\begin{subfigure}[c]{0.485\textwidth}
  \centering
  \scalebox{.9}{\includegraphics[width=\textwidth]{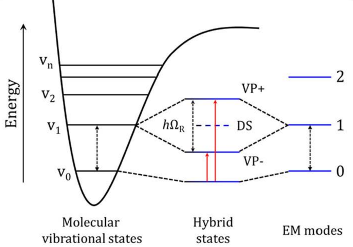}}
\end{subfigure}
\caption[An optical cavity]{Left panel: The experimental setup for a chemical reaction occurring inside of a Fabry--P\'erot optical cavity, adapted from Ref.~\cite{h2oSolvCoup}. Right panel: Coupling between molecular vibrations and cavity modes forms hybrid light-matter states known as the upper and lower vibrational polaritons (VP+/VP-) adapted from Ref.~\cite{Nagarajan2021} (right).}
\label{fig:opcav+hybrid.}
\end{figure}
\\
\indent 
When two highly reflective parallel mirrors are placed within micrometers of each other (shown in the left panel of Fig~\ref{fig:opcav+hybrid.}), photons - created by thermal quantum fluctuations of the vacuum field - resonate within them, forming standing waves in the enclosed electromagnetic field. These standing waves are known as \textbf{cavity modes}, and have frequencies determined simply by the spacing between said mirrors. For the $m^{\rm th}$ cavity mode, the frequency of the wave in the direction perpendicular to the mirrors of spacing $L_{\rm c}$ is
\begin{equation}
\omega_{\text{c},m} = \frac{mc\pi}{nL_{\rm c}} \text{,}
\end{equation}
where $n$ is the refractive index of the cavity and $c$ the speed of light \cite{Li2022}. If inside of this optical cavity there exist dipole-active molecular vibrational transitions of frequency close to $\omega_{c,m}$, they hybridise with the cavity mode, forming hybrid light-matter states known as \textbf{vibrational polaritons} (see right panel of Fig.~\ref{fig:opcav+hybrid.}). This combination is analogous to that observed in molecular orbital theory, such that the symmetric linear combination between photonic and vibrational states is stabilised relative to the unmixed states, and antisymmetric destabilised. These are known as the upper and lower polariton respectively.
\\ \indent 
The presence of these hybrid states is directly observable in the splitting of vibrational transition peaks in a cavity dipole absorption spectrum. The strength of interaction is quantified by the Rabi splitting observed for these peaks. When the cavity is resonant with an ensemble of $N$ vibrational transitions, this splitting is
\begin{equation}
    \Omega_{\rm R} = 2\sqrt{N}g_0 \text{,}
    \label{eqn:RabiSplitting}
\end{equation}
in which $g_0$ is the light-matter coupling strength, which will be further discussed in Sec~\ref{3:PFHam}. 
\FloatBarrier
\begin{figure}
\centering
\scalebox{0.85}{\includegraphics[width=\textwidth]{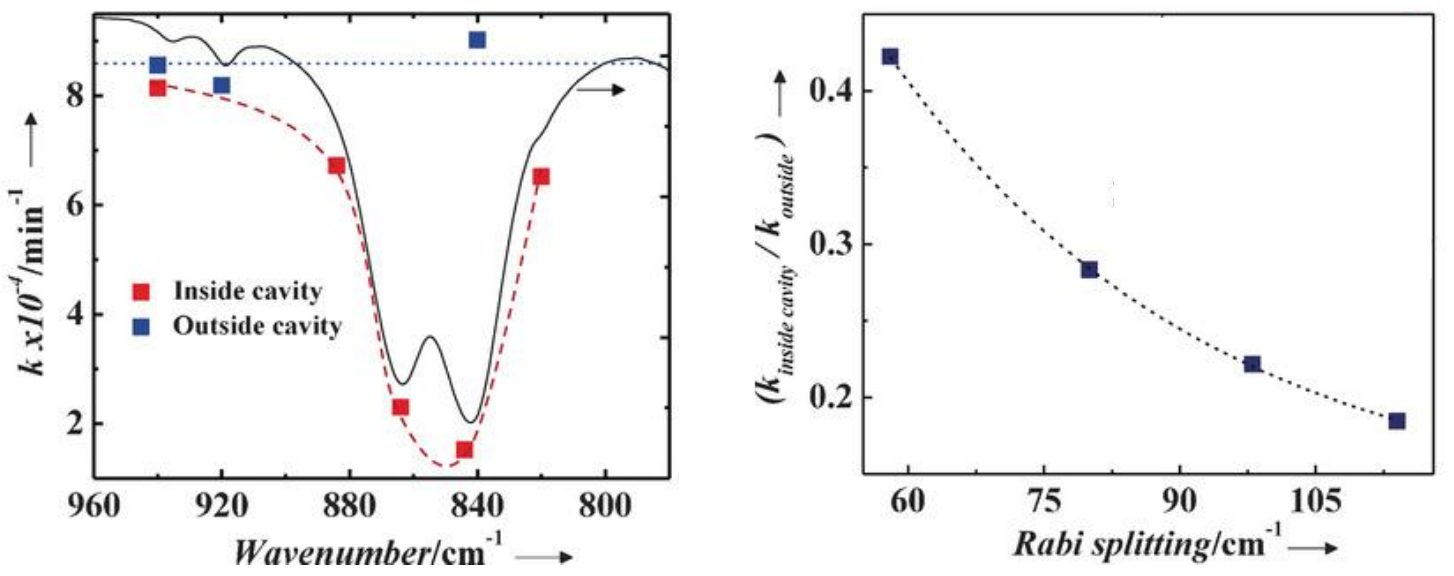}}
\caption[Experimental observations of resonance and the collective effect]{Cavity effects on the rate of the silane deprotection of 1-phenyl-2-trimethylsilylacetylene (PTA).  Left panel: The rate is maximally suppressed by a \textbf{factor of 5.5} when a cavity mode is resonant with a reactant infrared (IR) transition (black line) - known as the `Resonance Effect'. Right panel: Cavity rate effects increase with Rabi splitting - known as the `Collective Effect'. Adapted from Ref.~\cite{Thomas2016}}
\label{fig:expevid}
\end{figure}
It was first observed in 2012 \cite{1stratespaper} that performing certain chemical reactions inside of these optical cavities would significantly alter their rates. Since then, further experiments have found chemistry under vibrational strong coupling to influence chemoselectivity~\cite{ChemoCavity}, equilibrium constants~\cite{EqmCavity} and even stereoselectivity~\cite{WHCavity} of common chemical transformations. In addition, indirect modification of chemical reactivity via solvent coupling has also been observed \cite{h2oSolvCoup}, extending possible applications to biological systems, in which cavity coupling has been used to modify enzyme activity in aqueous media \cite{enzymeExp,enzymeExp2}.
\\
\indent 
Fig~\ref{fig:expevid} provides an example of this experimental evidence. 
In the left panel of Fig~\ref{fig:expevid} the rate of the silane deprotection of 1-phenyl-2-trimethylsilylacetylene (PTA) is suppressed by more than 5 times - but only if the optical cavity is resonant with the reactant's infrared active Si-C stretch (shown by the black line) - this is known as the \textbf{resonance effect}. 
The right panel of Fig~\ref{fig:expevid} shows that as the Rabi splitting (which is proportional to $\sqrt{N}$, as in Eq.~\ref{eqn:RabiSplitting}) is increased, the rate suppression of the deprotection reaction of PTA, when coupled to a resonant cavity mode is augmented. This is known as the \textbf{collective effect}.
\\
\indent
There is significant desire in the scientific community for a theoretical explanation of these effects, as it would allow experimentalists to more effectively harness this new tool in their transformations. Despite this motivation however, there is so far no widely accepted theoretical explanation for any of these findings - which is relatively unsurprising considering the magnitude of the effects observed the fact that the process involves no external input of energy. 
\\
\indent 
A common starting point is to consider a simplified model consisting of a single reaction vibrationally coupled to an optical cavity. In Ch.~\ref{3:} I will look for the resonance effect in a double-well model, comparing a variety of rate theories (introduced in Ch.~\ref{2:}) for a condensed-phase reaction.  
\\ 
\indent 
The experimental results found in the literature do not however refer to single reactions, but instead a large number (potentially as many as $\sim 10^{10}$) collectively coupled to the cavity. I will therefore in Ch.~\ref{4:} extend the analysis to an ensemble of $N$ reactants in order to understand cavity effects in the collective (experimentally relevant) regime. By explicitly simulating multiple reactive systems, I will examine the collective effect, as well as any changes in the resonance effect relative to the single molecule case.
\\
\indent
In Ch.~\ref{5:} I will then explore cavity effects on equilibrium constants, again looking for the resonance and collective effects. With these three approaches I hope to address all possible avenues by which chemical reaction dynamics may be modified by vibrational strong coupling, and perhaps provide some insight into the mechanism by which these effects operate.
\chapter{\label{2:}Rate Theory} 
\minitoc
In this chapter, I will introduce theories by which the unimolecular rates of reaction for both quantum and classical systems will be calculated. Assume that the configuration space of a molecular system can be separated into reactants A and products B, with chemical equation
\begin{equation}
% \underset{k_2}{\stackrel{k_1}{\rightleftharpoons}} FOR EQM SYMBOL
{\rm A}\underset{k_{\rm b}}{\stackrel{k_{\rm f}}{\dubarrow}} {\rm B} \text{.}
\end{equation}
For a general non-equilibrium initial condition, there will be an initial transient behaviour,\cite{lawrence2020a} which eventually settles to the kinetic equation such that
\begin{equation}
    \dot{P}_{\rm A}(t) = -k_{\rm f}\;P_{\rm A}(t) + k_{\rm b}\;P_{\rm B}(t)
\end{equation}
and 
\begin{equation}
    \dot{P}_{\rm B}(t) = k_{\rm f}\;P_{\rm A}(t) - k_{\rm b}\;P_{\rm B}(t) \text{,}
\end{equation}
in which $P_{\rm A}(t)$ and $P_{\rm B}(t)$ correspond to the population of A and B at time $t$ respectively.
The rate theories used yield $k_{\rm f}$, the rate of forward reaction. At equilibrium, the ratio of forward to backward reactions is equal to the equilibrium constant, hence the rate of backward reaction $k_{\rm b}$ can easily be inferred. In this thesis the potential surfaces used when discussing polariton effects on reaction rates will be symmetric with respect to the reaction coordinate, hence $k_{\rm f} = k_{\rm b} = k$.
\section{\label{2:CLT}Classical Rate Theory}
\indent The thermal rate constant for a reaction on a $1$-dimensional potential energy surface can be expressed as, 
\begin{equation}
     k\left(T \right) = \frac{1}{Q_{\rm r}} \lim_{t\to t_{\rm p}} c_{\rm fs}(t) \text{,}
\label{eqn:k-cfs}
\end{equation}\noindent
 where $Q_{\rm r}$ is the reactant partition function (per unit volume for reactions in three dimensions) \cite{Miller1983}. The dividing surface is defined as $q=q^\ddagger$. Here $c_{\rm fs}(t)$ is the correlation function between flux of reactants through the dividing surface at time $0$ and the population on the product side of this surface at $t$. The `plateau time' of $c_{\rm fs}$, $t_{\rm p}$ is is long enough such that the initial transient behaviour of the reactive system is removed from the calculation of rate. 
 % Note that equation ~\eqref{eqn:k-cfs} is valid as long as $c_{\rm fs}$ plateaus on a short timescale compared to the diminishing of the reactant - which is true for all systems discussed in this thesis. 
 The classical limit of $c_{\rm fs}$ is
\begin{equation}
    c_{\rm fs}^{\rm cl}(t) = \frac{1}{2\pi\hbar} \int dp_{0}\int dq_{0}\: e^{-\beta H(p_{0},q_{0})} \dot{\theta}(q_{0}-q^{\ddagger})\theta(q_{t}-q^{\ddagger}) \text{.}
\label{eqn:cfsCL}
\end{equation}
where $q_0$ and $q_t$ correspond to the reaction coordinate at time $0$ and $t$ respectively. This notation is also used for $p$, the momentum. The classical side operator, $\theta(q-q^\ddagger)$ is a Heaviside step function. The classical flux operator is its time derivative; 
\begin{equation}
    \dot{\theta}(q_0-q^\ddagger) \:=\:  \frac{d q_0}{d t}\delta(q_{0}-q^{\ddagger})
    \:=\:
    \frac{p_{0}}{m}
    \delta(q_{0}-q^{\ddagger})
     \text{ .}
\end{equation}
The integral in Eq.~\eqref{eqn:cfsCL} can therefore be seen as a sum going over only trajectories starting at the transition state, where $q_0 = q^{\ddagger}$ - such that $\delta(q_0 - q^\ddagger) \neq 0$. Each trajectory is weighted by its Boltzmann factor $e^{-\beta H(p_{0},q_{0})}$ and initial velocity $\frac{p_0}{m}$ - a weighting that therefore is positive if the trajectory's initial velocity is from reactants to products or negative if in the opposite direction.
At time $t$, all trajectories with $q_t \geq q^\ddagger$  will have a non-zero $\theta$ and will hence contribute their weight  to the value of $c_{\rm fs}(t)$. 
\\
\indent
If a given trajectory crosses the dividing surface at $t > 0$ it is considered to have `recrossed' the dividing surface. The effect of recrossing on $c_{\rm fs}(t)$ can be understood by considering two types of trajectory after some small time $\epsilon$ (before recrossing);
\begin{equation*}
\begin{split}
\text{i)  }&p_0 > 0 \implies q_{\epsilon} > q^\ddagger
\:\:\: \therefore \:\:\: \frac{p_0}{m}\theta(q_{\epsilon}-q^{\ddagger}) > 0  
\:\:\:\left[\text{positive contribution to } c_{\rm fs}(\epsilon)\right]
\\
\text{ii)  }&p_0 < 0 \implies q_{\epsilon} < q^\ddagger \:\:\:  \therefore \:\:\: \frac{p_0}{m}\theta(q_{\epsilon}-q^{\ddagger}) = 0 \:\:\:\left[\text{no contribution to } c_{\rm fs}(\epsilon)\right]
\end{split}
\end{equation*}
Now consider these scenarios at time $t$, after which recrossing - or any odd number of recrossings - has already occurred;
\begin{equation*}
\begin{split}
\text{i)  }&q_{t} > q^\ddagger
\:\:\: \therefore \:\:\: \frac{p_0}{m}\theta(q_{t}-q^{\ddagger}) = 0 
\:\:\:\left[\text{no contribution to } c_{\rm fs}(t)\right]
\\
\text{ ii)  }&q_{t} > q^\ddagger \:\:\: \therefore \:\:\: \frac{p_0}{m}\theta(q_{t}-q^{\ddagger}) < 0 \:\:\:\left[\text{negative contribution to } c_{\rm fs}(t)\right]
\end{split}
\end{equation*}
The result of recrossing is therefore either : i) - removal of a positive contribution or ii) - introduction of a negative contribution to $c_{\rm fs}(t)$. Hence as both types contribute negatively, $c_{\rm fs}^{\rm cl}$ is a decreasing function in time.
\begin{figure}[hbt!]
\centering
\includegraphics{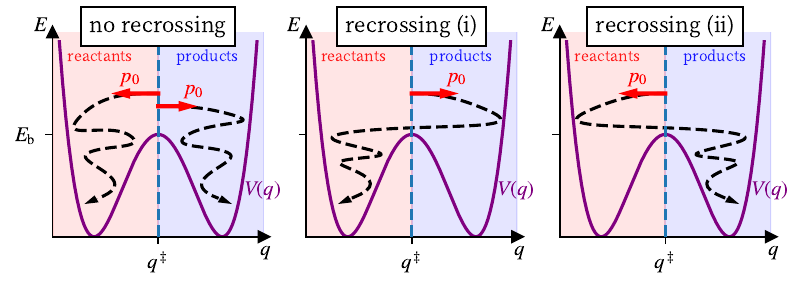}
\caption[Possible trajectory-based contributions to the flux--side correlation function]{A diagram showing possible trajectories' contributions to $c_{\rm fs}$}
\label{fig:D-1}
\end{figure}
\FloatBarrier
The full rate of reaction, including recrossing effects is therefore given by the plateau value of Eq.~\eqref{eqn:k-cfs}.
\section{\label{2:RPMD}Ring Polymer Molecular Dynamics (RPMD)}
\begin{figure}[hbt!]
\centering
\scalebox{.85}{\includegraphics[trim={1cm 0.35cm 0.35cm 0.35cm}]{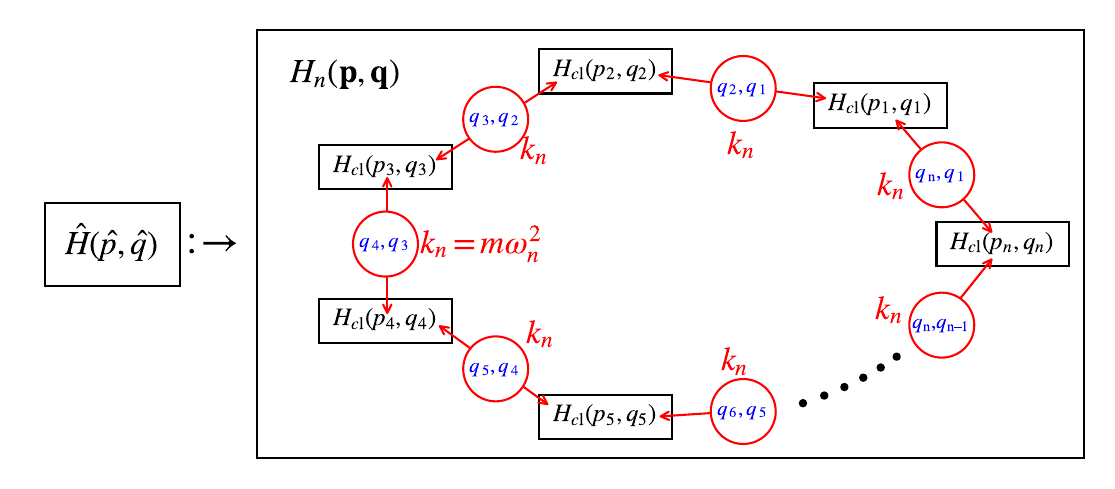}}
\caption[The ring polymer approximation to a quantum Hamiltonian]{A diagram showing the ring polymer approximation to  a quantum Hamiltonian. Blue text indicates which coordinates are linearly coupled together (i.e. coupling via harmonic springs with spring constant $k_n$). Red text indicates possible energy transfer pathways between beads (via said springs).}
\label{fig:D-RPMD}
\end{figure}

\indent Quantum rates of reaction could of course be obtained by substituting the quantum-mechanical flux-side correlation function, $c_{\rm fs}^{\rm qm}(t)$ into Eq.~\eqref{eqn:k-cfs}. In general - and especially in the condensed phase - simulation of this unfortunately would be prohibitively computationally expensive. Ring polymer molecular dynamics is a relatively inexpensive semi-classical method which offers a very good approximation to the quantum rate by using Feynman path integrals. In this section I will outline a simple derivation of the path integral form of a partition function \cite{Parrinello1984}, then explain how this can be extended such that it can apply to rate theory \cite{manocraig2004a}.
\\
\indent 
 In this derivation, integrals are over all space unless otherwise specified. The momenta and position coordinates, $p$ and $q$ are $1$-dimensional here, but multidimensional generalisation is straightforward (adding an index to each). The assumed information is as follows \cite{Dirac1927}; 
\[
\begin{split}
    {\rm tr}\left[ \hat{O}\right] \equiv \int dq \bra{q}\hat{O}\ket{q} 
    \: \: \: \: \: & \: \: \: \: \: 
    \bra{q}\ket{p} = \sqrt{\frac{1}{2\pi\hbar}}e^{\text{i}pq/\hbar}
    \\
    \hat{1} = \int dp \ket{p}\bra{p}
    \: \: \: \: \:& \: \: \: \: \: 
    \hat{1} = \int dq \ket{q}\bra{q} \text{.}
\end{split}
\]
\noindent The partition function of a system is the trace of the Boltzmann operator,
\begin{equation}
    Q = {\rm tr}\left[ e ^{-\beta \hat{H}}\right] \equiv 
    \int dq \bra{q} e ^{-\beta \hat{H}} \ket{q} \text{.}
\label{eqn:partfxn-statmech}
\end{equation}
Let us factorise Eq.~\eqref{eqn:partfxn-statmech} into $n$ terms as such,
\begin{equation}
    Q = \int dq \bra{q}\left(e^{-\beta_{n}\hat{H}}\right)   \left(e^{-\beta_{n}\hat{H}}\right) ... \ket{q} \text{,}
\end{equation}
where $\beta_{n} = \frac{\beta}{n} \equiv \frac{1}{n k_{\rm B}T}$, for which $n$ is a positive integer. Now, let us insert the identity in between each exponential ($n-1$ times). Each identity corresponds to integration over a different dummy variable, of which we will denote  $dq_1$ (corresponding to the first $dq$ in the trace above) $dq_2...dq_n$.
\begin{equation}
Q = \int dq_1 ... dq_n \bra{q_1} e^{-\beta_{n}\hat{H}}\ket{q_2}\bra{q_2}   e^{-\beta_{n}\hat{H}}\ket{q_3} ... 
    \bra{q_n}e^{-\beta_{n}\hat{H}}\ket{q_1} \text{.}
\label{eqn:factexpo}
\end{equation}
\noindent This integral now consists of $n$ factors of the form $\bra{q_{j}}\hat{O}\ket{q_{j+1}}$, with a cyclic index such that $q_{n+1}\equiv q_{1}$. Looking at each one of these terms, by considering the second order Taylor expansion of $e^{-\beta_{n}\hat{H}} = e^{-\beta_{n}(\hat{T}+\hat{V})}$ we get
\begin{equation}
    \bra{q_j}e^{-\beta_{n}\hat{H}}\ket{q_{j+1}} \simeq
    \bra{q_j}e^{-\frac{\beta_{n}\hat{V}}{2}}e^{-\beta_{n}\hat{T}}e^{-\frac{\beta_{n}\hat{V}}{2}}\ket{q_{j+1}} + O(1/n^3)\text{.}
\label{eqn:trotterapproxmn}
\end{equation}
\noindent The potential energies can now be evaluated, on `either side' (as $\hat{V}$ is hermitian), 
\begin{equation}
     \bra{q_j}e^{-\frac{\beta_{n}\hat{V}}{2}}e^{-\beta_{n}\hat{T}}e^{-\frac{\beta_{j}\hat{V}}{2}}\ket{q_{j+1}} =
     e^{-\frac{\beta_{n}V(q_j)}{2}}
     \bra{q_j}e^{-\beta_{n}\hat{T}}\ket{q_{j+1}}
    e^{-\frac{\beta_{n}V(q_{j+1})}{2}} 
    \text{.}
\end{equation}
Next, to evaluate $e^{-\beta_n\hat{T}}$ we insert the identity in the momentum eigenbasis, 
\begin{equation}
\begin{split}
       \bra{q_j}e^{-\frac{\beta_{n}\hat{V}}{2}}e^{-\beta_{n}\hat{T}}e^{-\frac{\beta_{j}\hat{V}}{2}}\ket{q_{j+1}} = & 
       \\
       \int \!\!dp&\:e^{-\frac{\beta_{n}V(q_j)}{2}}
     \bra{q_j}\ket{p}
     e^{-\frac{\beta_{n}p^2}{2m}}
     \bra{p}\ket{q_{j+1}}
    e^{-\frac{\beta_{n}V(q_{j+1})}{2}} \text{ .}
    \end{split}
\end{equation}
Inserting $\bra{q_j}\ket{p}$ and $\bra{p}\ket{q_{j+1}}$ and integration over $dp$ gives, 
\begin{equation}
\begin{split}
  \bra{q_j}e^{-\frac{\beta_{n}\hat{V}}{2}}e^{-\beta_{n}\hat{T}}e^{-\frac{\beta_{j}\hat{V}}{2}}\ket{q_{j+1}}    &  =  \int \!\!dp\: e^{-\frac{\beta_{n}(V(q_j)+V(q_{j+1})}{2}}
     e^{-\frac{\beta_{n}p^2}{2m}-\frac{-\text{i}p(q_{j}-q_{j+1})}{\hbar} }
     \\
      &  =  \frac{1}{2\pi\hbar}\sqrt{\frac{2\pi m}{\beta_n}} 
      e^{-\beta_n \left[ \frac{m\omega_n^{2}(q_j - q_{j+1})^2}{2}
      +\frac{V(q_j)}{2}
      +\frac{V(q_{j+1})}{2}\right]}
      \text{,}
\end{split}  
\end{equation}
where we have defined $\omega_n = \frac{1}{\beta_n \hbar}$. We can now re-introduce the integral over momenta. This is the key step that allows sampling of system using molecular dynamics (MD). Therefore,
\begin{equation}
    \bra{q_j}e^{-\beta_{n}\hat{H}}\ket{q_{j+1}} =
\frac{1}{2\pi\hbar}\int dp \:e^{-\beta_n \left[\frac{p^2}{2m} + \frac{m\omega_n^{2}(q_j -       q_{j+1})^2}{2}
      +\frac{V(q_j)}{2}
      +\frac{V(q_{j+1})}{2}\right]} + O(1/n^3) \text{.}
\end{equation}
This can now be re-inserted into Eq.~\eqref{eqn:factexpo}  with each integral over $dp$ getting its own index in the same way as $dq_1...dq_n$. Let us compact this notation such that $dq_1...dq_n \equiv \mathbf{dq}$, and analogously for momenta. Note that the overall error is now of order $1/n^2$ as the approximation of Eq.~\eqref{eqn:trotterapproxmn} is made $n$ times \cite{Craig2005}.
\begin{equation}
\begin{split}
    Q =&
\left(\frac{1}{2\pi\hbar}\right)^n
\int  \mathbf{dp}\int  \mathbf{dq}
  \prod_{j=1}^{n}
e^{\!\!\!-\beta_n\!\left[\!\frac{p_j^2}{2m} + \frac{m\omega_n^{2}(q_j - q_{j+1})^2}{2}
      +\frac{V(q_j)}{2}
      +\frac{V(q_{j+1})}{2}\!\right]} + O(1/n^2)
      \\
      \equiv&
      \left(\frac{1}{2\pi\hbar}\right)^n
\int  \mathbf{dp}\int  \mathbf{dq}
\:e^{-\beta_n 
\sum\limits_{j=1}^{n}\left[\frac{p_j^2}{2m} 
      + V(q_j)
      + \frac{m\omega_n^{2}(q_j -       q_{j+1})^2}{2}
      \right]} + O(1/n^2) \text{.}
\label{eqn:RPpartitionfxn}
\end{split}
\end{equation}
By inspection of Eq.~\eqref{eqn:RPpartitionfxn} one can see that the quantum partition function has been mapped that of a classical system in an extended phase space. This effective classical Hamiltonian is,
\begin{equation}
\begin{split}
    H_n(\mathbf{p},\mathbf{q}) = \sum\limits_{j=1}^{n}\left[
    \frac{p_j^2}{2m} + V(q_j)+ \frac{1}{2}m\omega_n^{2}(q_j - q_{j+1})^2\right] \text{, }
\label{eqn:RPMDham}
\end{split}
\end{equation}
which is comprised of $n$ copies of the system corresponding to the classical limit of the initial quantum Hamiltonian, $H_{\rm cl}(p_j,q_j)$, with adjacent copies linked together by classical harmonic springs, as illustrated in Fig.~\ref{fig:D-RPMD}.
The phase-space variables $\mathbf{p}$ and $\mathbf{q}$ are now vectors containing $p_j$ and $q_j$ respectively. The index $j$ corresponds to each 'bead' on the ring polymer necklace, with $n$ giving the total number of beads.
\FloatBarrier 
As $n \xrightarrow{}\infty$,  the error will tend to zero, but  for most systems of interest at ambient temperatures \cite{Lawrence2019} we find sufficient convergence with $n=32$ as this value makes $\frac{\beta\hbar\omega_{\rm max}}{n} \gg 1$, where $\omega_{\rm max}$ is the max frequency of said systems.
\subsection{Zero-Point Energy and Tunneling}
\indent RPMD goes beyond classical MD and includes quantum statistical effects such as zero-point energy (ZPE) and quantum tunneling. These effects can be very important in condensed phase reactions - in particular for systems including light atoms such as Hydrogen. At high temperature, the ring polymer springs become stiff and as such, the ring polymer shrinks to a classical particle. At low temperature however, the ring polymer 'swells' such that the bead distribution recovers the probability distribution of the quantum wave function. The swelling is about the \textbf{centroid}, the classical `position' of this quantum particle,
\begin{equation}
    \Bar{q} = \frac{1}{n}\sum_{j=1}^n q_j \text{.}
\label{eqn:Centroid}
\end{equation}
The thermally averaged radius of gyration of a free $n$ bead ring polymer is \cite{Lawrence2019}
\begin{equation}
    \Delta q_n = \sqrt{\frac{\beta\hbar^2}{12m}\left(1-\frac{1}{n^2}\right)} \text{ .}
\end{equation}
Note that swelling persists as $n \xrightarrow{}\infty$, and can be seen as a consequence of the Heisenberg uncertainty principle.
\\
\indent 
With this in mind, the potential experienced by the centroid relative to the potential averaged over the beads at some finite temperature can be calculated.
\begin{figure}[hbt!]
\centering
\includegraphics{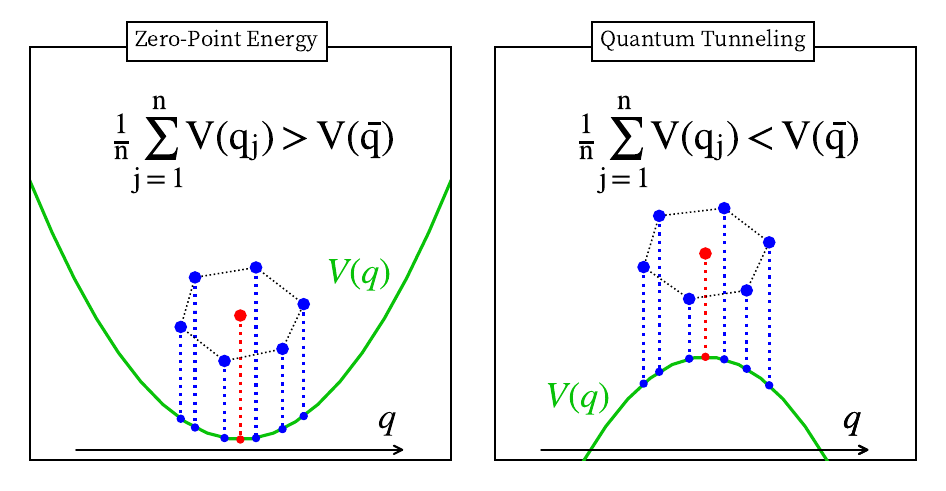}
\caption[RPMD captures zero-point energy and tunneling]{A diagram showing how the swelling of a ring polymer confined to a potential well (left) and passing over a potential barrier (right) captures zero-point energy and tunneling. Red dots annotate the centroid, and blue the ring polymer beads.}
\label{fig:D-zpe+tun}
\end{figure}
% \FloatBarrier
In a potential well (left panel of Fig.~\ref{fig:D-zpe+tun}) the potential averaged over beads is greater than that of the centroid. In this case, zero-point energy (ZPE) is added to the classical potential. When the ring polymer is passing over a barrier (right panel of Fig.~\ref{fig:D-zpe+tun}) however, the average potential experienced by the beads is smaller than that of the centroid. This implies that barrier crossing may occur even when the system energy is less than that of the classical barrier height, hence facilitating quantum tunneling through said barrier.
\\
\indent
Its derivation proves that RPMD will capture the exact quantum statistics of a system. It does not prove that real-time system dynamics are exact, and since it does not involve any phases that \textbf{RPMD cannot capture real-time quantum coherences}. The dynamics produced do conserve the Boltzmann distribution making RPMD nonetheless an effective method for approximating quantum correlation functions \cite{2013RPMDreview}. In addition, for reactions occurring in the condensed phase at room temperature, quantum decoherence is rapid - heavily reducing the coherences - making RPMD well suited for simulation of chemical reaction dynamics.
\section{\label{2:RPMDr.t.}RPMD Rate Theory}
\indent As RPMD is simply classical mechanics in an extended phase-space \cite{Craig2005},  one can define a function analogous to the classical flux-side correlation and use this to define a reaction rate.
\begin{equation}
    c_{\rm fs}^{\rm RPMD}(t) = \left(\frac{1}{2\pi\hbar}\right)^{n}
    \int \mathbf{dp}_0 \int \mathbf{dq}_0
    e^{-\beta_n H_n{(\mathbf{p}_0,\mathbf{q}_0)}}
    \Dot{\theta}(s(\mathbf{q}_0))
    \theta(s(\mathbf{q}_t)) \text{.}
\label{eqn:cfsRPMD}
\end{equation}
The dividing surface between reactants and products is defined at $s(\textbf{q})=0$. It can be shown \cite{Craig2005-II} that the rate of reaction is independent of the choice of $s(\textbf{q})$. A simple choice (which we will use in this thesis) is to use the centroid - as defined in Eq.~\eqref{eqn:Centroid} - of the ring polymer to define the dividing surface, such that
\begin{equation}
    s(\mathbf{q}) = \frac{1}{n}\sum_{j=1}^{n}\left(q_{j}\right) - q^\ddagger\text{.}
\end{equation}
The resulting flux operator $\dot{\theta}(s(\textbf{q}))$ is obtained by taking the time derivative of $\theta(s(\textbf{q}))$ with the help of the chain rule,
\begin{equation}
\begin{split}
    \dot{\theta}(s(\mathbf{q})) \:=&\: \delta(s(\mathbf{q}))\sum\limits_{j=1}^{n}\left[
    \frac{\partial s(\mathbf{q})}{\partial q_{j}} 
    \frac{d q_{j}}{d t}\right]
    \:
% \\
% =&\:
%     \delta(\bar{q})\frac{1}{n}\sum\limits_{j=1}^{n}
%     \frac{p_{j}}{m} \text{ .}
\\
=&\:
    \delta(\bar{q}-q^\ddagger)\frac{\bar{p}}{m} \text{,}
\end{split}
\end{equation}
where $\bar{p}$ is the centroid momentum, defined analogously to $\bar{q}$. Lastly, $H_n(\mathbf{p},\mathbf{q})$ is just as defined in Eq.~\eqref{eqn:RPMDham} and is formed by `ring polymerising' the classical limit of the quantum Hamiltonian.
\\
\indent
With this framework now in place, RPMD can be used to generate trajectories with which this integral in Eq.~\eqref{eqn:cfsRPMD} may be numerically evaluated. A set of initial phase points are generated, with the centroid fixed to the diving surface at $t=0$. Then, trajectories are numerically integrated in time according to Hamilton's equations:
\begin{equation}
    \frac{dq_{j}}{dt} = \frac{\partial H_n}{\partial p_{j}} \quad\text{,}\quad
    \frac{dp_{j}}{dt} = -\frac{\partial H_n}{\partial q_{j}} \text{ .}
\end{equation}
After each time step, $\theta(s(\mathbf{q}_t))$  measures if the trajectory is on the product side. The function $c_{\rm \:fs}^{\rm \:RPMD}(t)$ is calculated by averaging over many trajectories with associated weightings as mentioned in Sec.~\ref{2:CLT}, until convergence. The single bead limit of RPMD reduces to classical rate theory, as for $n=1$: $\Bar{q} \equiv q$ and $ H_{ n}(\mathbf{p},\mathbf{q})\equiv H_{\rm cl}(p,q)$, hence classical rates are obtained by the same method. Note that for increased stability in the integration of RPMD equations of motion we used the `Cayley propagator' - details of which can be found in Appendix~\ref{app:IneEOM}.
% These are obtained by application of the Euler-Lagrange equation onto phase space coordinates.\textbf{This} is why RPMD is not able to capture real-time quantum coherences, and as such will not be able to capture shallow tunneling as other methods such as HEOM  \cite{Lindoy2022} would be able to. The exact Quantum Lagrangian would be \cite{Hele2015} 
% \begin{equation}
%     \hat{L}_{QM} = \frac{p}{m} \frac{\partial}{\partial q} -
%     \sum_{\lambda = 1,odd}^{\infty} \frac{1}{\lambda !}
%     \left(\frac{i\hbar}{2}\right)^{\lambda -1}
%     \frac{\partial^{\lambda}V(q)}{\partial q^{\lambda}}
%     \frac{\partial^{\lambda}}{\partial p^{\lambda}}
% \end{equation}
% But instead RPMD uses the classical limit ($\hbar \xrightarrow{ } 0$) of this expression, corresponding to the classical Liouvillian, thus removing all terms of order $\hbar^2$ and above.
% \begin{equation}
%     L_{cl} = \frac{p}{m} \frac{\partial}{\partial q} -
%     \frac{\partial V(q)}{\partial q}
%     \frac{\partial}{\partial p}
% \label{eqn:LiouCl}
% \end{equation}
% With this framework, RPMD Rate theory can be used.
\section{\label{2:SBM}System--Bath Model}
\indent 
The effect of a solvent is often a crucial component in the understanding of chemical reaction dynamics. The system--bath model is commonly used to simulate the effects of a solvent, treating its impact on the reaction dynamics as an effective friction. To model this friction we couple our reaction to a bath of harmonic oscillators.
\begin{equation}
   \hat{H}(\hat{p},\hat{q},\vc{\hat{P}},\vc{\hat{Q}}) = \hat{H}_{\rm sys}(\hat{p},\hat{q}) + \hat{H}_{\rm bath}(\vc{\hat{P}},\vc{\hat{Q}},\hat{q}) \text{,}
\label{eqn:SysBathModel}
\end{equation}
where $ \hat{H}_{\rm sys}$ is the Hamiltonian of the isolated quantum system and
\begin{equation}
    \hat{H}_{\rm bath}(\vc{\hat{P}},\vc{\hat{Q}},\hat{q}) = 
    \sum_{i=1}^{n_{\rm B}}\left[
    \frac{\hat{P}_{i}^2}{2} + 
    \frac{1}{2}\omega_{\text{B}i}^{2}\left(\hat{Q}_{i} + \frac{c_{\text{B}i}}{\omega_{\text{B}i}^{2}}\hat{q}\right)^{2} \right] \text{.}
\label{eqn:Bath}
\end{equation}
The `bath' in question is simply $n_{\rm B}$ quantum harmonic oscillators (known as `bath modes') each of natural frequency $\omega_{\text{B}i}$. Each oscillator is linearly coupled to the system's reaction coordinate, $\hat{q}$, with coupling strength $c_{\text{B}i}$, facilitating reversible energy exchange between system and bath. These parameters are selected from the spectral density function,
\begin{equation}
    J(\omega) = \frac{\pi}{2} \sum_{i=1}^{n_{\rm B}} 
    \frac{c_{\text{B}i}^{2}}{\omega_{\text{B}i}}\delta(\omega-\omega_{\text{B}i}) \text{,}
\end{equation}
which contains all information about the bath and its coupling to the system. The spectral density is often modelled as a linear function at small $\omega$ multiplied by a cutoff function at large $\omega$. In this thesis we will use the Ohmic,
\begin{equation}
    J_{\rm Ohm.}(\omega) = \eta_{\text{B}}\omega e^{-\omega/\gamma_{\text{B}}}\text{,}
\end{equation}
and the Debye,
% \begin{equation}
%     J_{\rm Deb.}(\omega) = \eta_{\text{B}}\frac{\omega\:\gamma_{\text{B}}}{\omega^2 +\gamma_{\text{B}}^2}\text{,}
% \end{equation}
\begin{equation}
    J_{\rm Deb.}(\omega) = \eta_{\text{B}}\omega\frac{\gamma_{\text{B}}^2}{\omega^2 +\gamma_{\text{B}}^2}
\end{equation}
spectral densities, for both of which $\eta_\text{B}$ controls the magnitude of friction and $\gamma_\text{B}$ the shape of $J(\omega)$ of bath `B'.
\\
\indent 
To allow an explicit simulation only a finite number of bath `modes' - a harmonic oscillator of frequency $\omega_{\text{B}i}$ coupled with coupling $c_{\text{B}i}$ - are employed. Details of this discretisation and the parameters used can be found in Appendices~\ref{app:3-} and \ref{app:4-} respectively. 
\\
\\
\indent The notation used above will be continued throughout this thesis, with capitalised coordinates being used for the bath coupled to an un-capitalised coordinate just as the bath coordinates $\hat Q_i$ are coupled to the system coordinate $\hat q$ in Eq.~\eqref{eqn:Bath}.
\section{\label{2:TST}Transition State Theory (TST)}
\indent The Eyring equation \cite{Eyring1935} is a common tool used by the undergraduate chemist to obtain the Transition State Theory (TST) rate of reaction. For a uni-molecular reaction, 
\begin{equation}
 k^{\rm TST}(T) = \frac{1}{\beta h} \frac{Q_\ddagger}{Q_{\rm r}} e^{-\beta E_{\rm b}} \text{,}
 \label{eqn:Eyringeqn}
\end{equation}
where $Q_{\rm r}$ and $Q_\ddagger$ correspond to the reactant and transition state  partition functions respectively, $\beta = \frac{1}{k_{\rm B}T}$ where $T$ is the temperature and $k_{\rm B}$ the Boltzmann constant. As usual, $Q_\ddagger$ has a single vibrational degree of freedom removed, corresponding to movement over the barrier. Note that the zero-point energies of reactants and products are included within their partition functions, \textbf{not} $E_{\rm b}$ (which is the classical barrier height of the reaction).
\newline
\indent
It can be shown \cite{Craig2005-II} that the transition state theory rate of reaction can be obtained from $c_{\rm fs}^{\rm cl}(t\!\xrightarrow{}\!0^+)$ as
\begin{equation}
    k^{\rm TST} = \frac{1}{Q_{\rm r}}c_{\rm fs}(t\!\xrightarrow{}\!0^+) \text{,}
\end{equation}
where $c_{\rm fs}$ is either the classical or ring polymer flux-side correlation function, to yield the classical or quantum transition state theory rates respectively.
% At $t=0$ there is trivially no recrossing ...
% $k^{\rm HTST}$ is the Harmonic Transition State Theory (HTST) rate
\subsection{Harmonic Transition State Theory (HTST)}
A faster, but approximate, way of obtaining transition state theory rates is by making a harmonic approximation to the reactant and transitions state partition functions.
 \\
\indent Consider the partition function of a `ring polymerised' harmonic oscillator. Its $n$-bead Hamiltonian is
\begin{equation}
    H_n(\mathbf{p},\mathbf{q}) = \sum_{j=1}^{n} \left[H_{\rm cl}(p_j,q_j) + \frac{1}{2}m\omega_n^2(q_{j}-q_{j+1})^2\right]\text{,}
\end{equation}
where as usual $q_{n+1} \equiv q_1$ and
\begin{equation}
    H_{\rm cl}(p,q) = \frac{p^2}{2} + \frac{1}{2}\lambda^2q^2 
\end{equation}
is the Hamiltonian of a classical harmonic oscillator of natural frequency $\lambda$ with unit mass. It can be shown \cite{Craig2005} that the partition function of this system is 
 \begin{equation}
     Q_n = \prod_{j=1}^{n}
     \left(4\:\sin^2\left(\frac{j\pi}{n}\right) + \left(\frac{\beta\hbar\lambda}{n}\right)^2 \right)^{1/2}
     \text{.}
     \label{eqn:QrpHO}
 \end{equation}
 As was shown in Sec.~\ref{2:RPMD}, the $n\rightarrow\infty$ limit of Eq.~\eqref{eqn:QrpHO} will yield the exact partition function of a quantum harmonic oscillator. Evaluation of said limit was found in the appendix of \cite{Craig2005}.
\begin{equation}
     Q_{\rm qm} = \lim_{n\rightarrow \infty}\left[Q_n\right] = \frac{1}{2\:\sinh(\beta\hbar\lambda/2)}
     \text{ .}
\label{eqn:QHO-partitnfxn}
\end{equation}
This reduces to the classical partition function,
\begin{equation}
     Q_{\rm cl} \equiv Q_{n=1} = \frac{1}{\beta\hbar\lambda} \text{,}
\label{eqn:CHO-partitnfxn}
\end{equation}
when $\beta\hbar\lambda\ll1$ so the high temperature ($\beta\rightarrow0$) limit of Eq.~\eqref{eqn:QHO-partitnfxn} recovers the classical result in Eq.~\eqref{eqn:CHO-partitnfxn}, as expected.
\\
\indent
By taking a second  order Taylor expansion of $V(q)$, the potential energy surface along the reaction coordinate at either the reactants ($q=q_{\rm r}$) or transition state ($q=q^{\ddagger}$) stationary point, a local harmonic approximation of the Hamiltonian can be made;
\begin{equation}
    V_{\rm rct}(q) \simeq \frac{1}{2}m\omega_{\rm r}^2(q - q_{\rm r})^2
\:\:\:\:\:\text{and}\:\:\:\:\:
    V_{\ddagger}(q) \simeq E_{\rm b} - \frac{1}{2}m\omega_{\rm b}^2(q - q^{\ddagger})^2 \text{ .}
\end{equation}
With this expansion in place, all terms in the total classical Hamiltonian - even when including a dissipative bath - are of maximum order harmonic. This means that by taking normal modes - via diagonalisation of the mass-weighted Hessian, the Hamiltonian can be expressed as a sum of uncoupled harmonic oscillators, each with natural frequency $\lambda_i$.
\\
\\
\indent
The transition state Hamiltonian will have one imaginary frequency: $\text{i}\lambda_\ddagger$. This  frequency is discarded as it corresponds to movement over the reaction barrier; an inverted potential. The reactant and transition state partition functions, $Q_{\rm r}$ and $Q_\ddagger$ can then be constructed as a product of the partition functions for each normal mode either classical or quantum - and inserted into Eq.~\eqref{eqn:Eyringeqn}.
Therefore the classical rate constant is
\begin{equation}
    k^{\rm HTST}_{\rm cl} = \frac{1}{2\pi} e^{-\beta E_{\rm b}} \prod_i^\ddagger \frac{1}{\lambda_i} 
    \prod_j^r \lambda_j \text{ .}
\end{equation}
Its quantum counterpart is
 \begin{equation}
    k^{\rm HTST}_{\rm qm} =  \frac{2}{\beta h} e^{-\beta E_{\rm b}} \prod_i^\ddagger \frac{1}{\sinh(\beta\hbar\lambda_i/2)}
    \prod_j^r\sinh(\beta\hbar\lambda_j/2) \text{ ,}
\end{equation}
which now includes the contribution of zero--point energy to reaction rate, but will \textbf{not} include quantum tunneling through the barrier, as this degree of freedom  was removed.
\section{\label{2:KramTh}Kramer's Theory}
\begin{figure}[hbt!]
\centering
\includegraphics{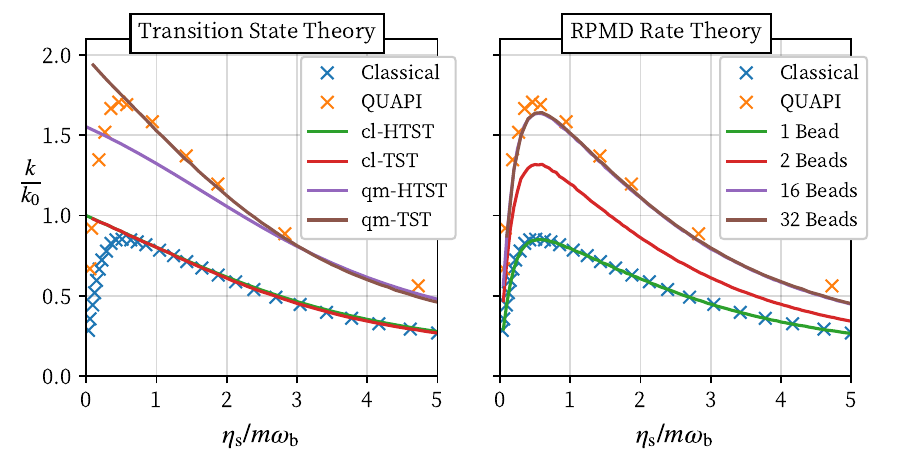}
\caption[A general example of Kramer's turnover]{Reaction rate dependence on solvent friction for the system of \cite{Topaler1994}, normalised by  $k_0$ the classical TST rate corresponding to $\eta_{\rm s} = 0$. The dimensionless ratio $\eta_{\rm s}/m\omega_{\rm b}$ measures the strength of the solvent friction relative to reactant mass $\times$ barrier frequency (a constant).} 
\label{fig:exampleKT}
\end{figure}

For both the classical and quantum rates, there exists a value of solvent friction at which the rate of reaction is maximised. This is known as \textbf{Kramer's Turnover} \cite{Kramers1940} and be explained in terms of the trajectory-based interpretation of $c_{\rm fs}$ as laid out in Sec.~\ref{2:CLT}, recalling that recrossing of the dividing surface of a trajectory strictly leads to a decrease in $c_{\rm fs}$.
\begin{figure}[hbt!]
\centering
\includegraphics{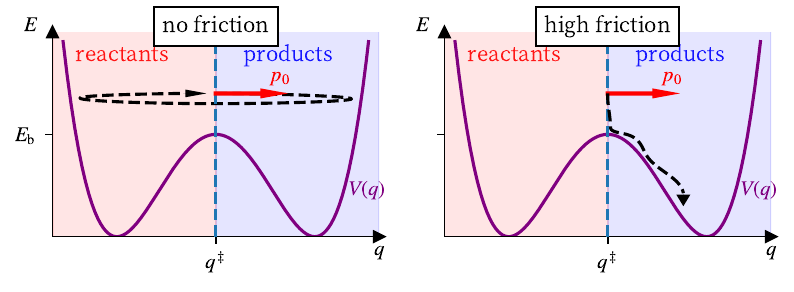}
\caption{Rate-limiting factors in high and low friction regimes}
\label{fig:kramtExplained}
\end{figure}
\FloatBarrier
For the limit of no solvent friction, no energy is dissipated into the surroundings. This means that all trajectories starting at the dividing surface cross from reactant to products, then recross to re-form reactants etc. As a consequence, the overall rate of reaction is zero, as there is no net flow from reactants to products (see left panel of Fig.~\ref{fig:kramtExplained}). As solvent friction is increased, the amount of recrossing decreases, thus increasing the rate of reaction. At the limit of high friction however there is no recrossing at all. The rate is instead decreased at high friction due to an increasingly small Boltzmann weighting (see Sec.~\ref{2:CLT}) being given to each trajectory. This effect is observed because the additional bath Hamiltonian (see Sec.~\ref{2:SBM}) is of course included in the total Hamiltonian by which this weighting is calculated.
Kramer's theory simply combines these limits giving a maximum in rate at some intermediate friction. This is known as `Kramer's Turnover'.
\\
\indent
With the aforementioned models in place, they can be tested against some exact quantum (Quasi-Adiabatic propagator Path Integral; QUAPI) and classical data \cite{Topaler1994} for a general double-well transition problem, shown in Fig.~\ref{fig:exampleKT}. Full details of the model are given in Ref.~\cite{Topaler1994}. Note that the vertical axis is the dimensionless ratio between $k$, the rate of reaction and $k_0$, the rate of classical transition state theory rate for $\eta_{\rm s} = 0$.
% \FloatBarrier
As transition state theory does not take into account recrossing of the dividing surface, it is not appropriate to calculate rates in the low-friction limit. After turnover however, it reproduces the exact results very well.  Quantum Harmonic Transition State Theory (qm-HTST) does not take into account tunneling through the barrier, and hence under-estimates the rate of reaction. Any discrepancy between classical Harmonic Transition State Theory (cl-HTST) and classical Transition State Theory (cl-TST) can be attributed to the anharmonicity of the barrier, but as can be seen it is very minor.
\\
\indent
The effect of recrossing can be taken into account with RPMD (right panel) with 1 and 32 beads almost quantitatively capturing the entire rate profile from low to high friction for classical and quantum rates respectively. As these methods are trajectory-based, there are errors in calculation that can be attributed to lack of convergence in: the length and number of trajectories, the number of bath modes on each bath, the number of ring polymer beads (for the quantum case) etc. For the rest of this thesis, 32 beads will be used for quantum RPMD simulations - which was found to give converged results in all cases. The details about other convergence parameters can be found in Appendix~\ref{app:4-}.
\chapter{\label{3:} A Single Reaction in a Cavity} 
\minitoc
\section{\label{3:1Pintro}Reaction Rates for a Symmetric Double Well}
\begin{figure}[hbt!]
\centering
\includegraphics[trim={0.25cm 0 0 0},clip,width=14cm,height=8cm]{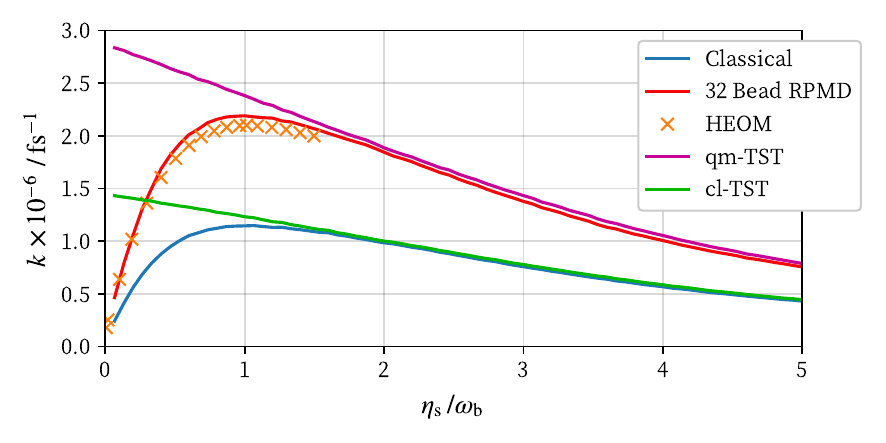}
\caption[Kramer's turnover for a single molecule outside of an optical cavity]{Kramer's turnover for a single molecule outside of an optical cavity. The dimensionless ratio $\eta_{\rm s}/\omega_{\rm b}$ is between the solvent friction strength and barrier frequency (a constant)}
\label{fig:KT-1mol.}
\end{figure}

In a recent paper \cite{Lindoy2022} Lindoy \textit{et al.\ }used Hierarchical Equations of Motion (HEOM), a numerically exact quantum method, to simulate reaction rates in a double-well potential. For a molecular system undergoing some generic reaction in a double--quartic potential well, the Hamiltonian is
\begin{equation}
\Hat{H}_{\rm mol} = \Hat{T}_{\rm mol}(\hat{p}) + \Hat{V}_{\rm mol}(\hat{q}) \text{,}   
\end{equation}
in which 
\begin{equation}
\Hat{T}_{\rm mol} = \frac{\hat{p}^2}{2}
\text{\:\:\:\:\:and\:\:\:\:\:}
\hat{V}_{\rm mol} = \frac{\omega_{\rm b}^4}{16E_{\rm b}}\hat{q}^{ 4} -\frac{\omega_{\rm b}^2}{2}\hat{q}^{ 2} \text{.}
\end{equation}
The parameters $\omega_{\rm b}$ and $E_{\rm b}$ give the classical barrier frequency and height relative to the bottom of the reactant well respectively. The reaction coordinate $\hat{q}$ is parametrically dependent on the molecular system's Cartesian coordinates, and $\hat{p}$ is the conjugate momentum. Note that we use mass-scaled coordinates throughout this thesis
\\
\indent
% Although an exact method, due to exponential scaling in basis and convergence issues HEOM can only simulate systems with a small number of states [HEOM CITATION] limiting therefore the size and shape of potentials used. 
Lindoy \textit{et al.\ }explored several parameter regimes some of which had a significant ground-state tunneling (GST) contribution to reaction rate. As this effect is untypical for reactions occurring in the condensed phase, the man focus in this thesis will be on a symmetric double well with $\omega_{\rm b}= \SI{500}{cm^{-1}}$ and $E_{\rm b} = \SI{2000}{cm^{-1}}$, for which the GST contribution to rate is negligible in comparison to the total rate. For this double well, the reactant harmonic frequency is $\omega_{\rm r} = \sqrt{2}\:\omega_{\rm b} =\SI{707}{cm^{-1}}$. The temperature for the simulations was set to $\SI{300}{K}$. 
\\
\indent
To simulate the effect of solvent friction, the system--bath model can be used. Hence, for a single reactant in an open system,
\begin{equation}
\hat{H}_{\rm tot} = \hat{H}_{\rm mol}(\hat{p},\hat{q})\: + \:\hat{H}_{\rm{solv}}(\vc{\hat{P}},\vc{\hat{Q}},\hat{q}) \text{,}
\end{equation}
where
\begin{equation}
 \hat{H}_{\rm{solv}} = \sum_{i=1}^{n_{\rm{S}}}\left[
    \frac{\hat{P}_{i}^2}{2} + 
    \frac{1}{2}\omega_{\text{S}i}^{2}\left(\hat{Q}_{i} + \frac{c_{\text{S}i}}{\omega_{\text{S}i}^{2}}\hat{q}\right)^{2}\right]
    \text{. }
\end{equation}
The parameter $\eta_{\rm s}$ is used to control the strength of coupling between the reaction coordinate and its environment (the solvent) as discussed in Sec.~\ref{2:SBM}. The values of discretised bath frequencies and coupling constants, $\{c_{\text{S}i}\}$ and $\{\omega_{\text{S}i}\}$, are  obtained from the solvent spectral density $J_{\rm S}(\omega)$ and $\eta_{\rm s}$ respectively as explained in Appendix~\ref{app:3-}.
\begin{figure}[hbt!]
\centering
\includegraphics[trim={0.25cm 0 0 0}]{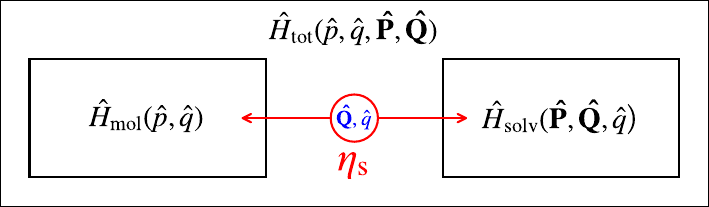}
\caption[The possible flow of energy for an open quantum system outside of a cavity]{A diagram to show the possible flow of energy for an open quantum system outside of a cavity. The red arrow indicates the flow of energy between sub-systems, blue text indicates which coordinates are linearly coupled together.}
\label{fig:D1}
\end{figure}
\FloatBarrier
\indent
First, we consider the reaction rate outside a cavity. The HEOM outside--cavity Kramer's Turnover behaviour is compared with that of the rate theories introduced in Ch.~\ref{2:} for this model in Fig.~\ref{fig:KT-1mol.}.
\FloatBarrier
\indent 
Classical rate theory does not include the effects of zero-point energy or quantum tunneling, both of which augment the reaction rate. For this reason, it underestimates the rate of reaction by roughly a factor of $2$. In contrast, RPMD rate theory includes these effects and can be seen to almost quantitatively capture the exact quantum Kramer's Turnover for this system.
By comparing to transition state theory, the effects of recrossing on the reaction rate can be seen to be significant for low friction and small for high friction, as expected from the discussion in Sec.~\ref{2:KramTh}.  
\section{\label{3:PFHam}The Pauli--Fierz Hamiltonian}
\indent
The Pauli--Fierz (PF) Hamiltonian is used to describe interactions between light and matter. Assumptions in its derivation involve the long wavelength approximation (assuming the cavity wavelength to be long relative to the extent of the system), the rotating wave approximation (neglecting rapidly oscillating terms in the Hamiltonian - very common in quantum optics) and the dipole approximation \cite{PnA}. For a molecule coupled to an optical cavity with a single optical mode of frequency $\omega_{\rm c}$,
\begin{equation}
\Hat{H}_{\rm PF} = \Hat{H}_{\rm mol}(\hat{p},\hat{q}) + \Hat{H}_{\rm cav}(\hat{p}_{\rm c},\hat{q}_{\rm c},\hat{\mu}) \text{  ,}
\label{eqn:PF}
\end{equation}
where 
\begin{equation}
\Hat{H}_{\rm cav} =  \frac{\hat{p}_{\rm c}^2}{2} + 
\frac{1}{2}\omega_{\rm c}^2 \left(\hat{q}_{\rm c} + \sqrt{\frac{2}{\hbar\omega_{\rm c}}}\eta_{\rm c}\hat{\mu}\right)^2  \text{,}
% \label{eqn:PF}
\end{equation}
which describes the optical cavity and its interaction with the system. Here, $\hat{\mu}(\hat q) = \vc{\hat e}\cdot\vc{\hat \mu}$ and is the dipole moment operator projected onto the electronic ground state, along the cavity polarisation $\vc{\hat{e}}$. In agreement with Lindoy \textit{et al.\ }we select the dipole moment to be linear in the molecular coordinate such that $\hat \mu =\hat q$. 
 % and $\hat{e}$ the unit vector corresponding to the polarisation direction of the cavity field
The system--cavity coupling constant is $\eta_{\rm c} =\frac{1}{\omega_{\rm c}} \sqrt{\frac{\hbar \omega_{\rm c}}{2 \epsilon_0 V  }}$, where $\epsilon_0$ is the system permittivity and $V$ is the system volume. For a Fabry--P\'erot optical cavity, $\omega_{\rm c} \propto V^{-1}$, hence making $|\eta_{\rm c}|$ independent of cavity frequency in this experimental setup. 
The Rabi splitting of the system of interest $\Omega_{\rm R}$ is directly proportional to $\eta_{\rm c}$. To keep the splitting on the order of $\SI{100}{cm^{-1}}$ we will use the same range as Lindoy \textit{et al.\ }with $1.25\times10^{-3}{\rm a.u.}\leq\eta_{\rm c}\leq5\times10^{-3}{\rm a.u.}$ (see Ref.~\cite{Lindoy2022}).
\begin{figure}[hbt!]
\centering
\includegraphics[trim={0.25cm 0 0 0}]{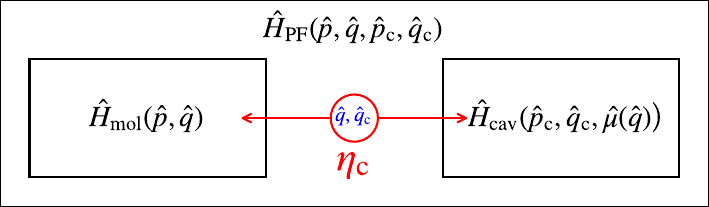}
\caption[The possible flow of energy for an isolated quantum system comprised of an optical cavity coupled to a molecule]{a diagram showing the possible flow of energy for an isolated quantum system comprised of an optical cavity coupled to a molecule. Red arrows indicate the flow of energy between sub-systems, blue text specifies which coordinates are linearly coupled together.}
\label{fig:DPF}
\end{figure}
\FloatBarrier
\indent
The position and momentum operators for the cavity field, $\hat{p}_{\rm c}$ and $\hat{q}_{\rm c}$ originate from the photon creation and annihilation operators $\hat{a}^\dagger$ and $\hat{a}$, which are the same as those of a harmonic oscillator, and hence can be mapped to position and momentum coordinates in the same way \cite{PnA}.
 \\
 \indent 
The cavity--system coupling in Eq.~\eqref{eqn:PF} has exactly the same form as in Eq.~\eqref{eqn:Bath}. This means that adding an optical cavity is essentially the same as coupling the system to an additional bath with $J(\omega) \propto \delta(\omega-\omega_{\rm c})$ - a single bath mode at $\omega = \omega_{\rm c}$, with a coupling strength $\propto \eta_{\rm c}$.

\section{\label{3:PR}Polariton Resonance}
The final addition required for a complete model is the cavity `loss' - pathways by which photons in the optical cavity are dissipated. Loss occurs due to either (i) leakage of energy due to imperfections in the mirror or (ii) interaction of the cavity with other charged particles in the surroundings.
\\
\indent 
Lindoy \textit{et al.\ }used the system-bath model to simulate these effects. They introduced the parameter $\tau_{\rm L}$, the \textbf{cavity lifetime}, with which the strength of cavity loss bath was inferred.
\begin{equation}
    \eta_{\rm L} = \frac{1}{\tau_{\rm L}}\frac{\omega_{\rm c}^2+\gamma_{\rm L}^2}{2\gamma_{\rm L}^2}
    \left(1- e^{-\beta\hbar\omega_{\rm c}}\right) \text{,}
\end{equation}
where $\gamma_{\rm L}$ is a parameter controlling the frequency distribution of the cavity loss, $J_{\rm L}(\omega)$, as discussed in Appendix \ref{app:3-}. Adding these effects together,
\begin{equation}
    \hat{H}_{\rm tot} = \hat{H}_{\rm mol}(\hat{p},\hat{q}) + 
    \hat{H}_{\rm{bath}}(\vc{\hat{P}},\vc{\hat{Q}},\hat{q}) + 
     \hat{H}_{\rm{cav}}(\hat{p}_{\rm{c}},\hat{q}_{\rm{c}},\hat{\mu}) +  
    \hat{H}_{\rm{loss}}(\vc{\hat{P}_{\rm{c}}},\vc{\hat{Q}_{\rm{c}}},\hat{q}_{\rm{c}})\text{,}
\label{eqn:1PfullHam}
\end{equation}
where  $\hat{H}_{\rm mol}$, $\hat{H}_{\rm{bath}}$ and  $\hat{H}_{\rm{cav}}$ are as in previous sections, and
\begin{equation}
    \hat{H}_{\rm{loss}} =   \sum_{i=1}^{n_{\rm{l}}}\left[\frac{\Hat{P}_{\text{c}i}^2}{2} +  \frac{1}{2}\omega_{\text{L}i}^{2}\left(\hat{Q}_{\text{c}i} + \frac{c_{\text{L}i}}{\omega_{\text{L}i}^{2}}\Hat{q}_c\right)^{2}\right]\text{.}
\label{eqn:LossHam}
\end{equation}
The notation used is the same as that of the previous sections;  $\{\omega_{\text{B}i}\}$ and $\{\omega_{\text{L}i}\}$ are the frequencies corresponding to the discretised system and cavity loss baths, with their coupling constants chosen from bath spectral densities $J_{\rm L}(\omega)$ and $J_{\rm B}(\omega)$ with a suitable discretisation protocol (see Appendix~\ref{app:3-}).
\begin{figure}[hbt!]
\centering
\includegraphics[trim={0.25cm 0 0 0}]{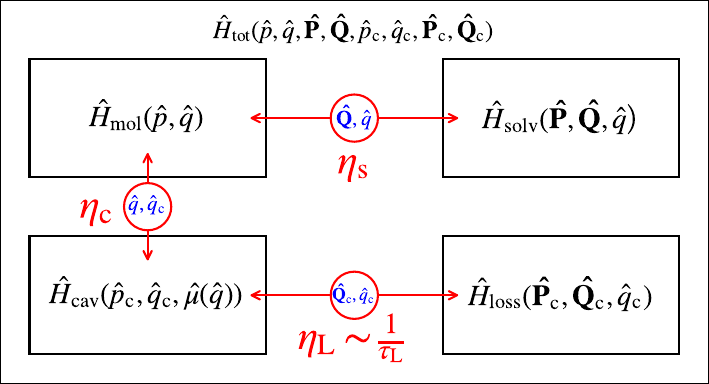}
\caption[The possible flow of energy for an open quantum system comprised of an imperfect optical cavity coupled to a single molecule]{a diagram to show the possible flow of energy for an open quantum system comprised of an imperfect optical cavity coupled to a single molecule. Red arrows indicate the flow of energy, blue text indicates which coordinates are linearly coupled together}
\label{fig:D2}
\end{figure}
\FloatBarrier
The dipole function $\hat{\mu}$ is set to equal $\hat{q}$ for the simulations as explained in Sec.~\ref{3:PFHam}. The RPMD Hamiltonian corresponding to $\hat{H}_{\rm tot}$ can be formed by `ring polymerising' it  (as introduced in Sec.~\ref{2:RPMD} and diagrammatically exemplified in Fig~\ref{fig:D-RPMD}). 
\FloatBarrier
\subsection{The Low Friction Regime}
\begin{figure}[!h]
\centering
\scalebox{.95}{
\includegraphics[trim={0cm .5cm 0.cm .25cm}]
{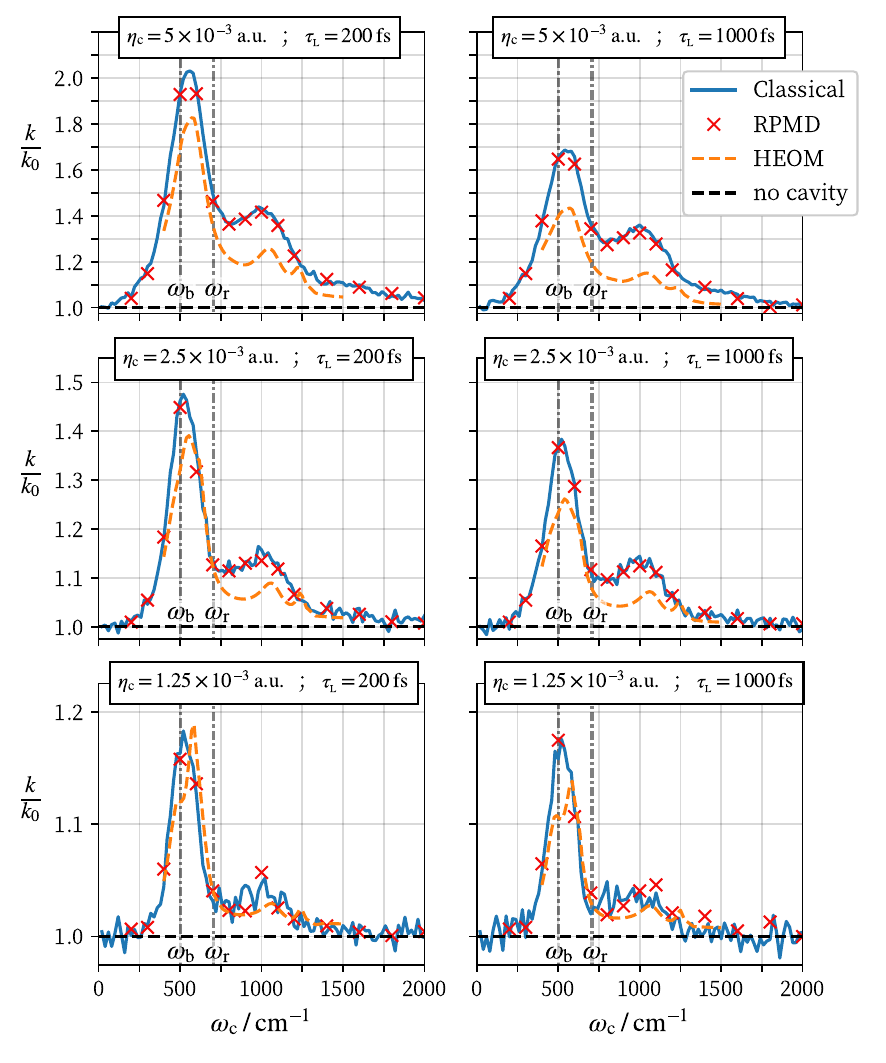}}
\caption[Cavity induced changes in rate in the low friction regime.]{Cavity induced changes in rate in the low friction regime. The rate $k$ is normalised by $k_0$, the corresponding rate outside the cavity, defined as that of the reaction with all parameters the same, but with $\eta_c = 0$.}
\label{fig:res_ED_1P}
\end{figure}

Setting ${\eta_\text{s}}/{\omega_{\text{b}}} = 0.1$ puts the system in the low friction regime, before turnover as shown in Fig.~\ref{fig:KT-1mol.}. 
Using the range of coupling strengths ($\eta_{\rm c}$) and loss timescales ($\tau_{\rm L}$) from Ref.~\cite{Lindoy2022}, let us explore the effect of a cavity on the rate of reaction, varying $\omega_\text{c}$. The effect is quantified by the dimensionless ratio, $k/k_0$ where $k$ is the rate of reaction and $k_0$ is the corresponding rate outside the cavity (defined as that of the reaction with all parameters the same, but with $\eta_c = 0$). These results are shown in Fig.~\ref{fig:res_ED_1P}.
% \FloatBarrier
\\
\indent
Somewhat surprisingly, Fig.~\ref{fig:res_ED_1P} shows that RPMD and classical mechanics give almost exactly the same results (to graphical accuracy) for polariton rate enhancement. This will be discussed in Sec.~\ref{3:recros.}.
\\
\indent
HEOM yields several large peaks which are captured to some extent by both 1 and 32 Bead RPMD. The peak positions do not coincide with that of the reactant frequency $\omega_{\rm r}$ (which is typically the peak position observed in experiments), but are in agreement over all three methods. The peak heights, on the other hand, are found to be slightly overestimated by classical/RPMD rates compared to the HEOM results. Sec.~\ref{3:recros.} argues that the RPMD and classical cavity effects $k/k_0$ are the same because all of the cavity effect is in the recrossing dynamics - of which are shown to be identical in Fig.~\ref{fig:TransCoef1P}. Since this recrossing dynamics - as quantified by $c_{\rm fs}(t)$ - is a highly nonlinear function of $\omega_{\rm c}$, simple explanations of peak widths and heights in terms of `resonance' (such as those provided in experimental papers) are not supported by calculations.
\\
\indent
Increased cavity loss, (signified by \textbf{decreased} $\tau_\text{L}$) is found to enhance the rate of reaction. In the supplementary information of their HEOM paper, Lindoy \textit{et al.\ }explored this effect, and found that the classical rate was relatively insensitive to this compared to the quantum rate. Increased cavity loss will lead to increased energy dissipation of the system via energy flow into the cavity and fast thermalisation with the loss bath (see Fig.~\ref{fig:D2}). In this low friction regime, the increased rate of energy dissipation will increase the rate as expected by Kramer's Theory.
\\
\indent
HEOM's peak heights decrease in magnitude at a slower rate as $\eta_{\rm c}$ is decreased, and are more effected by cavity loss than the other rate theories used. This suggests that some aspect of coherent recrossing effects \cite{Topaler1994} is contributing to the rate, which cannot be captured by RPMD (or classical MD).
\subsection{The High Friction Regime}
\begin{figure}[!ht]
\centering
\includegraphics[trim={0.25cm 0.25cm 0.cm 0.25cm}]
{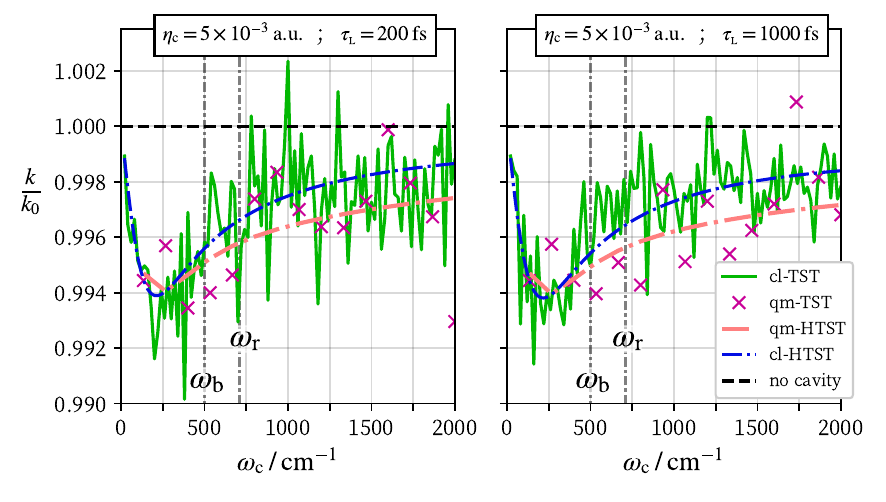}
\caption[Cavity induced changes in rate in the high friction regime]{Cavity induced changes in rate in the high friction regime.The rate $k$ is normalised by $k_0$, the corresponding rate outside the cavity, defined as that of the reaction with all parameters the same, but with $\eta_c = 0$.}
\label{fig:res_SD_1P}
\end{figure}

\indent
In the high friction regime, transition state theory is a suitable method to calculate rates due to the almost complete suppression of recrossing.
\\
\indent 
Multiple groups have investigated the cavity rate problem using various transition state theories, including Yang and Cao who used Harmonic Transition State Theory (HTST) \cite{Yang2021}. They considered a simpler Hamiltonian without solvent interactions or cavity loss;
\begin{equation}
\begin{split}
 \hat{H}_{\rm Cao} = \hat{H}_{\text{mol}} + \hat{H}_{\text{cav}}
  = \left[\frac{\Hat{p}^2}{2} + \hat{V}(\hat{q})\right]
+\left[\frac{\Hat{p}_{\rm c}^2}{2} + \frac{1}{2}\omega_{\rm c}^2(\hat{q}_{\rm c} + g\hat{q})^2\right] \text{.}
\label{eqn:1pCaoH}
\end{split}
\end{equation}
The unitless light-matter coupling constant, $g \propto \sqrt{\frac{1}{\omega_{\rm c}}}$, giving the same scaling dependence of coupling with cavity frequency as in the previous model. For this reason, Eq.~\eqref{eqn:1pCaoH} corresponds to the $\eta_{\rm s} = \eta_{\rm L} = 0$ limit of the model depicted in Fig.~\ref{fig:D2}. 
They found that the cavity only leads to a small $\sim\!1\%$ decrease in reaction rate over a broad range of $\omega_c$, with no indication of `resonance' with either $\omega_{\rm b}$ or $\omega_{\rm r}$.
% \FloatBarrier
\\
\indent
In Fig.~\ref{fig:res_SD_1P}, we investigate if these results change when explicitly including solvent friction and cavity loss.
The exact same parameters as in the previous section are used, but now with $\eta_\text{s}/\omega_{\rm b} = 4\:$ to explore polariton rate effects in the high friction regime. The effect of the cavity on the rate is found to be tiny compared to the low-friction case. For this reason, despite using many ($2\times10^5$) samples for transition state theory, the statistical errors of the full TST results (both classical and quantum) are almost as large as the effect itself. Nevertheless, the results indicate that there is no substantial difference between full TST and HTST, which is free of statistical errors. In either case, the effect of the cavity is small enough to say that the model does not explain experimental findings.
\\
\indent
Also note that the rate suppression occurs over a broad range of frequencies rather than at a sharp resonance frequency. In other words, Cao's results for the simpler model are essentially unchanged by including solvent friction and cavity loss.
\subsection{\label{3:recros.}Recrossing Effects on Reaction Rates}
\indent
The results of Fig.~\ref{fig:res_ED_1P} and Fig.~\ref{fig:res_SD_1P} imply that the main contribution to cavity-induced reaction rate adjustment can be understood entirely in terms of changes in recrossing of the dividing surface. This is quantified by the recrossing coefficient,
\begin{equation}
    \kappa = k/k_{\rm TST}\text{,}
\end{equation}
the ratio between the flux-side rate (which includes recrossing) and the transition state theory rate (which does not). 
\begin{figure}[!ht]
\centering
\scalebox{.85}{
\includegraphics[trim={0cm 0cm 0cm 0cm}] %LFT BTM RGT TOP%
{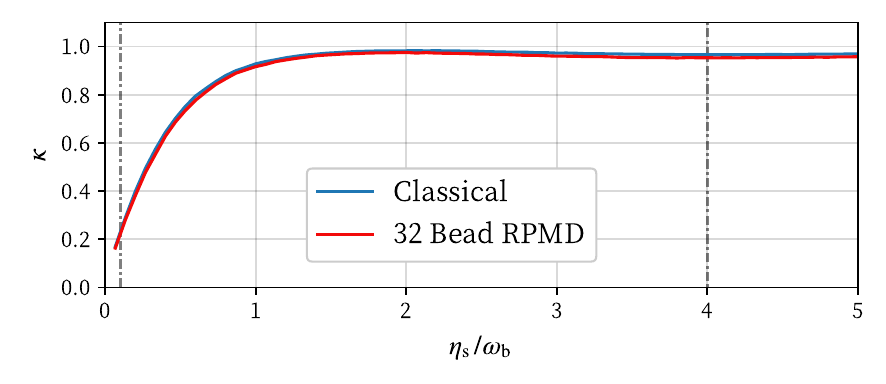}}
\caption[Friction dependence of the transmission coefficient $\kappa$ in classical dynamics and RPMD]{Friction dependence of $\kappa$ for classical mechanics and RPMD, corresponding to the Kramer's Turnover graph in Fig.~\ref{fig:KT-1mol.}. Dotted lines refer to the friction strengths used for the two regimes in Sec.~\ref{3:PR}.}
\label{fig:TransCoef1P}
\end{figure}
% \FloatBarrier
\noindent
Addition of an optical cavity to the system can be seen as addition of an extra bath mode, as discussed in Sec.~\ref{3:PFHam}. As a consequence, the solvent friction is effectively increased when the cavity is added. This increases $\kappa$ when in the low-friction regime by the same amount (as shown in Fig.~\ref{fig:TransCoef1P}) for \textbf{both} classical and RPMD - explaining why the rate enhancement is exactly the same, as shown in Fig.~\ref{fig:res_ED_1P} for which $\eta_{\rm S} = 0.1\omega_{\rm b}$. Similarly, the cavity has little effect on the rate for $\eta_{\rm S} = 4\omega_{\rm b}$ in Fig.~\ref{fig:res_SD_1P}, as in this regime the cavity has a negligible effect on the amount of recrossing.

\section{\label{3:Conclusion}Conclusions}
In this chapter, we have explored polariton effects on single-molecule reactions. We have found:
\begin{enumerate}
\item The RPMD and classical MD cavity effects in Fig.~\ref{fig:res_ED_1P} (the low solvent friction regime) are similar to the exact quantum (HEOM) results for a large range of system-cavity coupling strengths and cavity loss timescales, but slightly more pronounced. 
\item The quantum and classical transition state theory cavity effects in Fig.~\ref{fig:res_SD_1P} (the high friction regime in which TST is a valid approximation - see Sec.~\ref{2:KramTh}) are negligible, consistent with Yang and Cao's conclusion for their simpler model \cite{Yang2021}.
\item The RPMD and classical MD cavity effects in Fig.~\ref{fig:res_ED_1P} are the same because the vast majority of the cavity effect is in the recrossing dynamics, which are the same for RPMD and classical MD as shown in Fig.~\ref{fig:TransCoef1P}. They are presumably more pronounced than the cavity effect in HEOM because of coherent effects in the HEOM recrossing dynamics (which RPMD and classical MD do not capture).
\item There is no evidence in \textbf{any} of these results for a `resonant' cavity effect on the rate constant (as observed experimentally).
\end{enumerate}
These conclusions of course only apply to a single reaction occurring in a single mode optical cavity, under the Hamiltonian set out in Eq.~\eqref{eqn:1PfullHam}. As cavity the effect on reaction rate at high solvent friction is insignificant relative to that at low friction, the focus in the next chapter will lie solely on the low friction regime, to see if these relatively significant effects persist as the number of molecules collectively coupled to the optical cavity increases.
\chapter{\label{4:}Collective Effects on Reaction Rates} 
\minitoc
\section{\label{4:G2N}Generalisation to $N$ Reactants}
\begin{figure}[!ht]
\centering
\includegraphics[trim={0.25cm 0.25cm 0.cm 0.25cm}]
{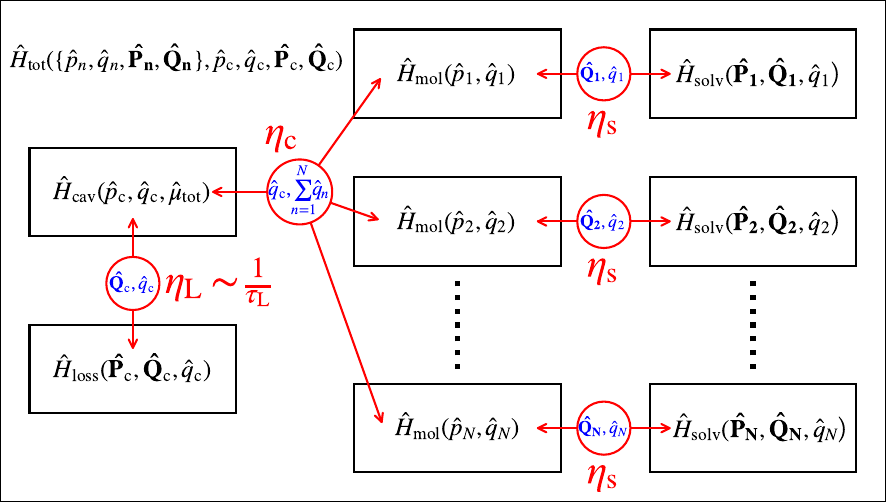}
    \caption[The possible flow of energy for an open quantum system comprised of $N$ molecules inside of a lossy optical cavity]{A model of the possible flow of energy for an open quantum system comprised of $N$ molecules inside of a lossy optical cavity. Red arrows indicate the flow of energy, blue text indicates which coordinates are linearly coupled together.}
\label{fig:D-NP}
\end{figure}

In this generalisation, let us assume that molecules are well separated inside of an optical cavity such that each reactant's interactions with the surroundings may be assumed to be independent. With this in mind, Eq.~\eqref{eqn:1PfullHam} can be generalised to that of $N$ reactants by summing over the individual molecule + bath Hamiltonians, and coupling the cavity to the \textbf{total} dipole of the system;
\begin{equation}
\begin{split}
    \hat{H}_{\rm tot} = \sum_{n=1}^N \left[\hat{H}_{\text{mol}}(\hat{p}_n,\hat{q}_n) + 
    \hat{H}_{\text{bath}}(\vc{\hat{P}_n},\vc{\hat{Q}_n},\hat{q}_n)\right]& 
    \\
    + \:\: \hat{H}_{\text{cav}}(\hat{p}_{\rm{c}},&\hat{q}_{\rm{c}},\hat{\mu}_{\rm tot}) \:+\:  
    \hat{H}_{\text{loss}}(\vc{\hat{P}_{\rm{c}}},\vc{\hat{Q}_{\rm{c}}},\hat{q}_{\rm{c}})
\label{eqn:NPfullHam}
\end{split}
\end{equation}
The dipole of the system, $\hat{\mu}_{\rm tot}=\sum\limits_{n=1}^N \hat{\mu}(\hat q_n)=\sum\limits_{n=1}^N \hat{q}_n$ , in analogy with the previous section. Fig.~\ref{fig:D-NP} shows the structure of this Hamiltonian diagrammatically.
\FloatBarrier

\section{\label{4:CaoArguement}Transition State Difficulties}
\begin{figure}[!h]
\centering
\includegraphics[trim={0.25cm 0.25cm 0.cm 0.25cm}]
{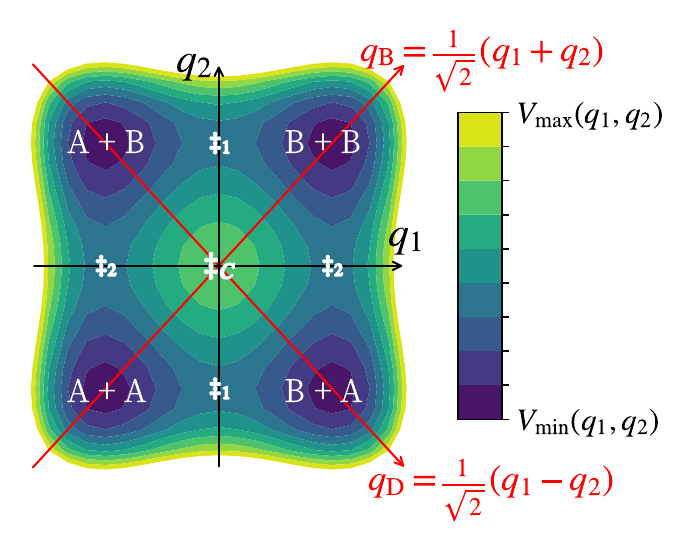}
    \caption{`Collective excitation' requires $N$ times larger activation energy and gives incorrect saddle-point index for `transition state'}
\label{fig:cao}
\end{figure}

Yang and Cao \cite{Yang2021} have considered polariton effects on quantum rates in the collective regime. They reported a collective scaling up of polariton effect observed with centroid transition state theory, with the caveat of this requiring the polariton to be `activated collectively to the transition state'. In their argument, the resulting potential for $N$ particles at the `collective transition state', denoted by $\ddagger_{\rm C}$, was approximated by a second order Taylor expansion in the normal mode basis of the molecular coordinate system;
\begin{equation}
    \left.V_{\rm N}(\{q_{n}\})\right|_{\ddagger_{\rm C}} \:\simeq E_{\rm a} - \frac{1}{2}\omega_{\rm b}^2 q_{\rm B}^2 + \sum _{i=1}^{N-1}\frac{1}{2}\omega_{\rm r}^2q_{\text{D}i}^2 \text{  .}
\end{equation}
Here, $q_{\text{D}i}$ is the $i$-th `dark state' (each of the $N - 1$ possible non--totally--symmetric linear combinations of the $N$ reactant coordinates, which are uncoupled from the optical cavity; hence `dark'), oscillating at the single molecule reactant frequency $\omega_{\rm r}$. The barrier curvature, $\omega_{\rm b}$ corresponds to that of the single molecule system, and $E_{\rm a}$ the activation energy required for this reaction pathway. The `bright state at the barrier'
\begin{equation}
    q_{\rm B} = \frac{1}{\sqrt{N}}\sum_{n=1}^N q_{n}  \text{,}
\end{equation}
the totally symmetric linear combination of mass-scaled reaction coordinates of the $N$ identical reactants. This expansion of $V(q)$ is troublesome however, as can be explained by considering the simplest case, $N = 2$. The potential energy surface for this 2-molecule system - in the absence of an optical cavity is
\begin{equation}
 V(q_1,q_2) = V_{\rm mol}(q_1) +V_{\rm mol}(q_2)   \text{ ,}
\end{equation}
where $V_{\rm mol}(q)$ is a 1--dimensional double-well potential as described in Sec.~\ref{3:1Pintro}, with transition state and reactant/product minima located at $q=0$ and $q=\pm q_{\rm r}$ respectively. This surface is visualised in Fig~\ref{fig:cao}
\FloatBarrier
\noindent
The potential maximum (which would correspond to the transition state) for the collective reaction coordinate $q_{\rm B}$ is located at $\ddagger_{\rm C}$. The activation energy required to reach this point is $E_{\rm a} = V_{\rm mol}(0) +V_{\rm mol}(0) = 2E_{\rm b}$. Due to its exponentially smaller Boltzmann weighting, this pathway would have a negligible contribution to the reaction rate relative to `non-coherent' reactions occurring via $\ddagger_1$ or $\ddagger_2$ with activation energy $E_{\rm a} = V_{\rm mol}(q_{\rm r}) +V_{\rm mol}(0) = E_{\rm b}$ when uncoupled to the optical cavity.
In addition, if one is to say
\begin{equation}
    \left.\left(\frac{\partial^2 V(q_1,q_2)}{\partial q_{\rm B}^2}\right)\right|_{\ddagger_{\rm C}} = -\omega_{\rm b}^2 \text{  ,}
\end{equation}
then by symmetry (as $V_{\rm mol}(q)$ is the same function for each coordinate)
\begin{equation}
    \left.\left(\frac{\partial^2 V(q_1,q_2)}{\partial q_{\rm D}^2}\right)\right|_{\ddagger_{\rm C}} = -\omega_{\rm b}^2 \text{    , not   } \omega_{\rm r}^2\text{ ,}
\end{equation}
which would make this `transition state' a maximum in the potential rather than a saddle point, making no physical sense. This reasoning can be extended to $N$ reactants, in which `collective excitation to the transition state' is at a maximum point in the potential, with $N$ times the activation energy. For this reason, most if not all reactions should go via `non-coherent' reaction pathways which are first order saddle points in the potential, with single-molecule activation energies.
\\
\indent 
In the same paper it was shown that the cavity effects on `non-coherent' reactions decreased to zero as the number of reactants coupled to the optical cavity was increased. This therefore suggests that transition state theory is not be able to capture the `collective effect' as observed experimentally, further justifying the decision to perform simulations in the low friction regime, where recrossing effects are most pronounced.
\section{\label{4:DS-N}Direct Simulation of $N$ Reactants}
The ambiguities in the definition of a dividing surface for an $N$-molecule reactant system coupled to a cavity can be removed by simply \textbf{not imposing a collective dividing surface}. Instead, one could set up a direct simulation by sampling the Boltzmann distribution of the $N$ reactants, then evolving and measuring the amount of product formed as a function of time. The rate of reaction can then be inferred by fitting the uni-molecular rate constant to the population curve. This can be derived by considering the rate process
\begin{equation}
% \underset{k_2}{\stackrel{k_1}{\rightleftharpoons}} FOR EQM SYMBOL
{\rm A}\underset{k_{\rm b}}{\stackrel{k_{\rm f}}{\dubarrow}} {\rm B} \text{.}
\end{equation}
As the rate of reaction considered here corresponds to that of a symmetric double well, by detailed balance, $k_{\rm f} = k_{\rm b} = k$. For initial population conditions subject to the population conservation constraint, $P_{\rm A}(t) + P_{\rm B}(t) = 1$,
\begin{equation}
\begin{split}
\dot{P}_{\rm A} =k(1 - 2P_{\rm A}) \text{ .}
\end{split}
\end{equation}
This can be integrated, followed by applying the initial conditions  $P_{\rm A}(0) = 1$, to yield 
\begin{equation}
{P}_{\rm A}(t) = \frac{1}{2}\left(1+{\rm e}^{-2kt}\right)\text{,}
\end{equation}
which can be rearranged to 
\begin{equation}
-\frac{1}{2}\ln\left(2{P}_{\rm A}(t) - 1\right) = kt\text{,}
\end{equation}
which allows the rate of reaction to be inferred via linear regression. The initial non-rate-like behaviour was found to occur for at most the first $\SI{500}{fs}$ of simulation and was therefore discarded in all data. 
\begin{figure}[!h]
\centering
\includegraphics[trim={0.25cm 0.5cm 0.cm 0.25cm}]
{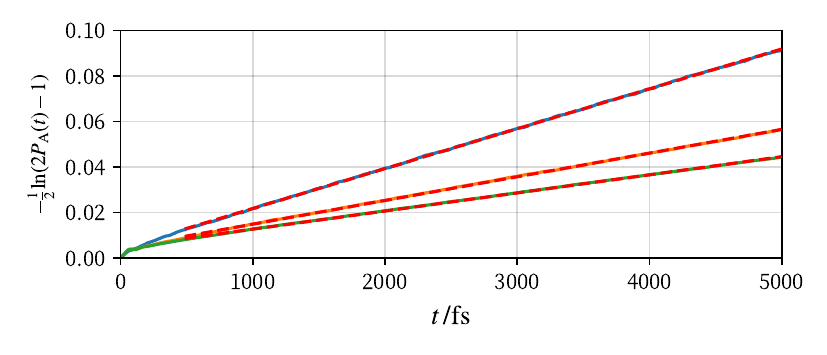}
\caption[Fitting procedure for direct simulation]{Procedure by which rate constants are obtained from population decay in direct simulation. Coloured lines represent a couple of examples of simulated data used to make Fig.~\ref{fig:ResNP}, dotted red lines are linear fits used to obtain $k$.}
\label{fig:DSfitting}
\end{figure}
\FloatBarrier
\noindent
In this way $k$ can be fitted to the population curve obtained by direct simulation, to give the rate of reaction for $N$ reactants coupled to an optical cavity without the need for any arbitrary definition of  a collective dividing surface.
\subsection{\label{4:themodel}The Model}
The simulation was divided into two stages. As the results obtained with RPMD in Ch.~\ref{3:} were very similar to classical mechanics, the methods used were purely classical (for computational ease).
\\
\\
\indent Stage 1: Equilibrium sampling of reactant configurations. This was done by a molecular dynamics simulation with the Hamiltonian of Eq.~\eqref{eqn:NPfullHam}, but with all trajectories constrained such that barrier crossing was prohibited. This was achieved by altering the potential energy surface in $H_{\rm mol}$ such that
\begin{equation}
    V_{\rm mol}(q)\leftarrow
    \begin{cases}
        \:\:V_{\rm mol}(q) \:\:\left(= \frac{\omega_{\rm b}^4}{16E_{\rm b}}q^{ 4} -\frac{\omega_{\rm b}^2}{2}q^{ 2} \right)& \text{if } q \leq 0\\
        \:\:\infty & \text{if } q > 0\\
    \end{cases}
\end{equation}
This effectively installed an elastic barrier at the dividing surface, $q = 0$, `bouncing' all barrier-crossing trajectories back into the reactant well. At regular interval of \SI{100}{fs} (to get uncorrelated samples), the coordinates and momenta of all particles were recorded and stored.
\\
\\
\indent Stage 2: Simulation of reactant trajectories. Each sample taken in the previous stage was then used as a set of initial conditions for an unconstrained simulation (i.e. with no infinite barrier in at $q=0$). Molecular dynamics was then used to progress these conditions in time. After each time-step the fraction of molecules on each side of the dividing surface was measured. Explicitly, this defines the population of reactants as 
\begin{equation}
    p_{\rm A}(t) = \frac{1}{N_{\rm s}}\sum_{s=1}^{N_{\rm s}}\left(\frac{1}{N}
    \sum_{n=1}^N \theta\left(-q_{n}^{(s)}(t)\right)\right)
\end{equation}
and the population of products as
\begin{equation}
        p_{\rm B}(t) = \frac{1}{N_{\rm s}}\sum_{s=1}^{N_{\rm s}}\left(\frac{1}{N}
    \sum_{n=1}^N \theta\left(q_{n}^{(s)}(t)\right)\right) \text{,}
\end{equation}
where $\theta(x)$ is a Heaviside step function, $N_{\rm s}$ is the number of samples obtained in stage 1 and $N$ the number of reactants coupled to the cavity as specified in the Hamiltonian. In contrast with the single-molecule studies, in which the simulation temperature was set to \SI{300}{K}, the temperature was \textbf{elevated to 500\,K} for these direct simulations. This was done to make barrier crossing more frequent and as such allow both shorter simulations with fewer samples.
\\
\indent 
To minimise the computational cost of simulation of many degrees of freedom (recall that each molecule is connected to its own solvent bath), the system-bath dynamics were integrated using the generalised Langevin equations (GLE) (see Appendix~\ref{app:3-}). These effectively replaced each bath with a single degree of freedom which evolved stochastically and thereby mimicked the effect of friction on the system.
\subsection{Polariton Resonance in the Collective Regime}
As mentioned in the introduction, the Rabi splitting of an infrared band in resonance with a cavity mode serves as a simple measurement of the light-matter coupling,
\begin{equation}
    \Omega_{\rm R} \propto \:\eta_{\rm c}\sqrt{N} \text{.}
\end{equation}
To isolate the effect of changing the number of molecules for a constant Rabi splitting, we scale down $\eta_{\rm c}$ by $\sqrt{N}$ for an ensemble of molecules to keep $\eta_{\rm c}\sqrt{N}$ constant. As $N\to\infty$, the coupling constant will of course tend $0$. The important question is though: as one approaches the collective regime where $N$ is on the order of $10^{10}$, will the cavity effects persist? 
\\
\indent
To answer this question the top left panel of Fig.~\ref{fig:res_ED_1P} was reproduced, but now with the temperature at $\SI{500}{K}$ (and all other parameters kept exactly the same). The same reaction was then simulated using the procedure outlined in Sec~\ref{4:themodel}, followed by subsequent simulations with an increased number of molecules, for a constant Rabi splitting. 
\begin{figure}[!h]
\centering
\includegraphics[trim={0.25cm 0.25cm 0.cm 0.25cm}]
{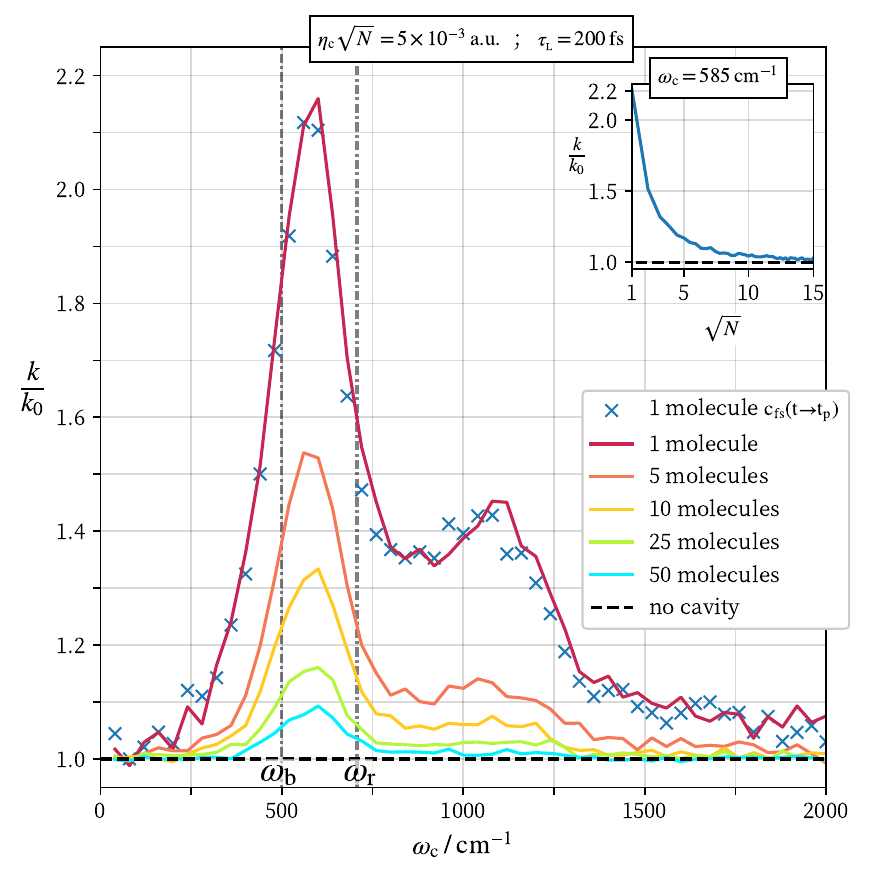}
\caption[Collective polariton resonance in low-friction regime]{Collective polariton resonance in low-friction regime (${\eta_\text{s}}/\omega_{\text{b}} = 0.1$) at an elevated temperature of $\SI{500}{K}$. Solid lines denote results obtained using direct simulation. Crosses are used for flux-side correlation results. The inset figure shows the maximal effect on reaction rates - found to be for $\omega_{\rm c} = \SI{585}{cm^{-1}}$ - decreasing to zero over an even larger range of $N$.}
\label{fig:ResNP}
\end{figure}
\FloatBarrier
\indent
First, we confirm that classical rate theory ($c_{\rm fs}(t\rightarrow t_{\rm p})$) agrees almost perfectly with the direct simulation, shown in Fig.~\ref{fig:ResNP} by the overall agreement between blue crosses and red line.
As expected, we find that the effect of the cavity on each system decreases as the $N$ is increased, for a constant Rabi splitting. The peak positions seem to be unchanged throughout. The inset figure shows the decrease in maximum peak height with an even larger range of N, implying that as $N$ approaches experimentally realistic values, the cavity rate enhancement will be negligible. If it is assumed that all relevant effects are captured to some extent by the methods used here, then the Hamiltonian must be insufficient in its description of chemical reactions occurring in an optical cavity. This is a startling conclusion, considering that it is widely used in literature \cite{li_cavity_2020,Mandal2020,Mandal2022}.
\section{\label{4:NP-Conclusions}Conclusions}
In retrospect, it is relatively unsurprising that the optical cavity's effect on a single molecule's properties vanishes as $N$ increases. This is due to the fact that the optical cavity couples to the \textbf{total} dipole of the system.
\begin{equation}
    \hat \mu_{\rm tot} = \sum_{n=1}^{N}\hat\mu(\hat q_{n}) \text{,}
\end{equation}
which under the Pauli--Fierz Hamiltonian of \ref{3:PFHam} gives a light-matter interaction potential
\begin{equation}
 \Hat{V}_{\rm cav}(q_{\rm c},\mu_{\rm tot}) = 
\frac{1}{2}\omega_{\rm c}^2 \left(\hat q_{\rm c} + \sqrt{\frac{2}{\omega_{\rm c}}}\eta_{\rm c}\sum_{n=1}^{N}\hat \mu(\hat q_{n})\right)^2  \text{.}   
\end{equation}
The cavity is therefore only coupled to one linear (the totally symmetric) combination of the dipoles of the individual reactants. This is also shown diagrammatically in Fig.~\ref{fig:D-NP}. In other words, the interaction strength per molecule is reduced by a factor of $\sqrt{N}$ for a constant Rabi splitting for $N$ molecules, causing all cavity effects to be `scaled out' in the collective regime. 
\\
\indent The cavity affects the bulk infrared absorption spectrum of the system as a whole, but as the number of degrees of freedom is increased, the effect on each individual molecule decreases, as shown by Fig.~\ref{fig:ResNP}, and therefore with this Hamiltonian the introduction of an optical cavity will not affect any single-molecule properties of a bulk system. This section has used a classical theory, but due to the fact that HEOM found the same qualitative behaviour as shown in Fig.~\ref{fig:res_ED_1P}, I expect the scaling-out argument to apply also in full quantum mechanics. This chapter concludes the exploration into polariton effects on chemical reaction rates. The next chapter discusses possible polariton effects on equilibria, attacking the problem from a more theoretical standpoint, now using quantum statistical mechanics.

% Motivation - Setup a QM system with a minimal basis set to allow test of this 'scaling-out' of effects observed in Classical Direct Simulation and TST.

\chapter{\label{5:}Cavity Effects on Equilibrium Constants} 
\minitoc
The equilibrium constant of the charge-transfer complexation between mesitylene and iodine, when occurring within an optical cavity, has been reported \cite{EqmCavity} to either double or decrease by four times simply depending on which infrared active vibrational degree of freedom the cavity was in resonance with. Similar results have been reported for benzene and iodine. To explore these effects let us consider a unimolecular equilibrium process,
\begin{equation}
  {\rm A} \xrightleftharpoons[k_{\rm b}]{\,\,\,k_{\rm f}\,\,\,} {\rm B} \text{.}
\end{equation}
The equilibrium constant is defined as
\begin{equation}
    K = \frac{k_{\rm f}}{k_{\rm b}} = \frac{Q_{\rm B}}{Q_{\rm A}}\text{,}
\end{equation}
where $Q_{\rm A}$ and $Q_{\rm B}$ correspond to the molecular partition functions of A and B. This ratio can also be expressed in terms of probabilities, 
\begin{equation}
    K = \frac{p_{\rm B}}{p_{\rm A}}\text{,}
\label{eqn:eqmPROB}
\end{equation}
where $p_{\rm A}$ and $p_{\rm B}$ are the probabilities of being in state A and B. This chapter will explore cavity effects on equilibria, firstly from a theoretical standpoint, followed by a simulation.
\section{\label{5:theory1P}Single-Molecule Effects}
First, we define $p_{\rm A}$ and $p_{\rm B}$ in statistical mechanics. Consider a single molecular system in the absence of an optical cavity. For a dividing surface located at $q=0$, the reactant (A) projection operator is
\begin{equation}
    \hat \theta_{\rm A} = \theta(-\hat q) \text{,}
\end{equation}
and product (B) operator is
\begin{equation}
    \hat \theta_{\rm B} = \theta(\hat q) \text{,}
\end{equation}
where $\theta(x)$ is the Heaviside step function. As the reaction can yield either A or B,
\begin{equation}
    \hat \theta_{\rm A} + \hat \theta_{\rm B} = \1 \text{.}
\label{eqn:heavisideIdentity}
\end{equation}
 For a quantum system comprised of a single molecular sub-system coupled to an optical cavity, the product basis states are
\begin{equation}
    \ket{q_{\rm c}\:q} = \ket{q_{\rm c}}\otimes\ket{q} \text{.}
\end{equation}
In the combined cavity-molecule system, the reactant projector is
\begin{equation}
    \hat h_{\rm A} = \1 \otimes \hat \theta_{\rm A}  \text{,}
\end{equation}
with $\hat h_{\rm B}$ defined analogously. Again, $\hat h_{\rm A} + \hat h_{\rm B} = \1\otimes\1$. The probability of A can therefore be expressed as a trace over the entire molecule-cavity space as
\begin{equation}
    p_{\rm A} = \frac{\Tr{e^{-\beta \hat H}\hat h_{\rm A}}}{Q} \text{,}
\label{eqn:pA1p}
\end{equation}
where the total partition function of the system is
\begin{equation}
   Q =  \Tr{e^{-\beta \hat H}} \text{,}
\end{equation}
and $\hat H$ is the total cavity-molecule Hamiltonian as defined in Eq.~\eqref{eqn:PF}. The probability of B is defined analogously. From Eq.~\eqref{eqn:eqmPROB}, the equilibrium constant for this reaction is therefore 
\begin{equation}
    K = \frac{\Tr{e^{-\beta \hat H}\hat h_{\rm B}}}{\Tr{e^{-\beta \hat H}\hat h_{\rm A}}} \text{.}
\end{equation}
Focusing our attention to the numerator, 
\begin{equation}
    \Tr{e^{-\beta \hat H}\hat h_{\rm A}} =  \intinf dq_{\rm c}\intinf dq \bra{q_{c}\:q} e^{-\beta \hat H}\hat h_{\rm A} \ket{q_{c}\:q} \text{.}
\end{equation}
By applying the Heaviside step function, the limits of this integral are changed such that
\begin{equation}
     \Tr{e^{-\beta \hat H}\hat h_{\rm A}} =  \intinf dq_{\rm c}\int_{\!-\infty}^{\:0}\!\! dq \bra{q_{c}\:q} e^{-\beta \hat H}\ket{q_{c}\:q} \text{.}
    \label{eqn:const.int}
\end{equation}
At room temperature, $k_{\rm B}T \approx \SI{200}{cm^{-1}}$. For the double-well model of the past chapters, the barrier height is $E_{\rm b} = \SI{2000}{cm^{-1}}$, meaning that the position distribution in A is heavily localised about the reactant potential minimum. Near this minimum, the potential is locally harmonic. Hence at ambient temperatures, the position distribution can be approximated as Gaussian such that,
\begin{equation}
     \Tr{e^{-\beta \hat H}\hat h_{\rm A}} \approx  \intinf dq_{\rm c}\intinf dq \bra{q_{c}\:q} e^{-\beta \left(\hat H_{\rm A}+\hat H_{\rm cavA}\right)} \ket{q_{c}\:q} \text{,}
\label{eqn:harmaproxeq1}
\end{equation}
where
\begin{equation}
    \hat H_{\rm A} = \frac{\hat p^2}{2} + E_{\rm A} + \frac{1}{2}\omega_{\rm A}^2\hat q^2  \text{,}
    \label{eqn:HarmHA}
\end{equation}
in which $E_{\rm A} + \frac{1}{2}\omega_{\rm A}^2\hat q^2$ is the second order Taylor expansion of $\hat V(\hat q)$ about the reactant minimum. Note that we have returned the upper limit of the integral over $q$ in Eq.~\eqref{eqn:harmaproxeq1} to infinity. This is valid because once we have made the harmonic approximation to the reactant potential in \eqref{eqn:HarmHA} the Boltzmann probability distribution in $q$ will be negligible for $q>0$. The approximation~\eqref{eqn:harmaproxeq1} improves as $\beta E_{\rm b} \to\infty$, due to increased localisation of this Boltzmann position distribution. Note also that this simple one-dimensional model can easily be generalised to multiple degrees of freedom by adding an index to $\hat q$ and replacing $\omega_{\rm A}^2\hat q^2$ by the appropriate quadratic form involving the Hessian of the multidimensional potential surface. Just as in previous sections,
\begin{equation}
   \Hat{H}_{\rm cavA}(\hat{p}_{\rm c},\hat{q}_{\rm c},\hat \mu(\hat q)) =
   \frac{\hat p_{\rm c}^2}{2} +\frac{1}{2}\omega_{\rm c}^2\left(\hat q_{\rm c} + g_{\rm A}\hat q\right)^2 \text{,}
\end{equation}
in which the constant $g_{\rm A}$ is the effective light--matter coupling constant at the reactant minimum. To perform the integral~\eqref{eqn:harmaproxeq1}, we diagonalise
\begin{equation}
    \mt{V^{\prime\prime}_{\rm A}} = \left(\begin{matrix}
    \omega_{\rm c}^2  & g_{\rm A}\omega_c^2 \\
    g_{\rm A}\omega_{\rm c}^2 & \omega_{\rm A}^2 + g_{\rm A}^2\omega_{\rm c}^2
    \end{matrix}\right) \text{,}
\end{equation}
the Hessian matrix of $\hat H_{\rm A} + \hat H_{\rm cavA}$. The eigenvalues of this matrix are 
\begin{equation}
    \Omega^2_{\rm A\pm} = \frac{1}{2}\left( \omega_{\rm A}^2 + g_{\rm A}^2\omega_{\rm c}^2+\omega_{\rm c}^2\right) \pm \frac{1}{2}\sqrt{\left( \omega_{\rm A}^2 + g_{\rm A}^2\omega_{\rm c}^2-\omega_{\rm c}^2\right)^2 + 4g_{\rm A}^2 \omega_{c}^4} \text{.}
\label{eqn:1pEigvals}
\end{equation}
As the diagonalisation is a unitary transformation, the integral~\eqref{eqn:harmaproxeq1} can be transformed with a unit Jacobian, allowing it to be expressed in terms of the product of partition functions of uncoupled harmonic oscillators of frequencies $\Omega_{\rm A\pm}$, with a constant energy shift $E_{\rm A}$. The same is of course true for B.
\\
\indent 
The quantum-mechanical equilibrium constant for a polariton system, using Eq.~\eqref{eqn:QHO-partitnfxn} is therefore
\begin{equation}
    K = \frac{\sinh\left(\beta \hbar \Omega_{\rm A-}/2\right)\sinh\left(\beta \hbar \Omega_{\rm A+}/2\right)}
    {\sinh\left(\beta \hbar \Omega_{\rm B-}/2\right)\sinh\left(\beta \hbar \Omega_{\rm B+}/2\right)}
    e^{-\beta(E_{\rm B}-E_{\rm A})} \text{.}
\label{eqn:K1pQM}
\end{equation}
Hence, for a (quantum) single molecular system, an optical cavity \textbf{can} effect equilibrium constants. This is true though only if either $\omega_{\rm A} \neq \omega_{\rm B}$ or $g_{\rm A}\neq g_{\rm B}$. Otherwise we would have that $\Omega_{A\pm} = \Omega_{B\pm}$, and the cavity effect would cancel. 
\\
\indent
Taking the classical $\beta\to0$ limit of~\eqref{eqn:K1pQM} gives
\begin{equation}
    K^{\rm cl} = \frac{\Omega_{\rm A+}\Omega_{\rm A-}}
    {\Omega_{\rm B+}\Omega_{\rm B-}}
    e^{-\beta(E_{\rm B}-E_{\rm A})} \text{,}
\end{equation}
which, after inserting Eq.~\eqref{eqn:1pEigvals} for each $\Omega$, simplifies to 
\begin{equation}
    K^{\rm cl} = \frac{\omega_{\rm A}}
    {\omega_{\rm B}}
    e^{-\beta(E_{\rm B}-E_{\rm A})} \text{,}
\label{eqn:K1pCL}
\end{equation}
which is the equilibrium constant of the reaction uncoupled from the optical cavity. This shows that while an optical cavity may effect quantum-mechanical equilibrium constants, it will have no effect classically.
The main question though is will the effect on the quantum equilibrium constant survive in the collective limit?
\section{\label{5:theoryNP}$N$-Molecule Effects}
For a system containing $N$ reactions, the probability that a given reaction selected at random is in A is given by the projection operator
\begin{equation}
    \hat h_{\rm A} = \frac{1}{N}\sum_{i=1}^N \hat h_{\text{A}i}\text{,}
\end{equation}
which is constructed from the projection operator of each subsystem into state A, such that $\hat h_{\rm A} + \hat h_{\rm B} = \1$ just as for the $N=1$ case. Using statistical mechanics, the probability of A is therefore written as
\begin{equation}
    p_{\rm A} = \frac{\Tr{e^{-\beta \hat H}\hat h_{\rm A}}}{Q} \text{.}
\label{eqn:pANp}
\end{equation}For a product basis defined by
\begin{equation}
    \ket{q_{\rm c}\:q_{\rm 1}\:q_{\rm 2}\:...\:q_{\rm N}}
    =
    \ket{q_{\rm c}}\otimes \ket{q_{\rm 1}}\otimes \ket{q_{\rm 2}}\:...\otimes \ket{q_{\rm N}} \text{,}
\end{equation}
these projection operators 
\begin{equation}
   \hat h_{\text{A}i} = \1 \otimes\underbrace{\1 \otimes ...\otimes \1}_{i-1 \text{ factors}} 
   \otimes\: \hat \theta_{\rm A} \otimes 
   \underbrace{\1 \otimes ... \otimes \1}_{N-i \text{ factors}} \text{,}
\end{equation}
with $\hat \theta_{\rm A}$ defined exactly as in the previous section. Eq.~\eqref{eqn:heavisideIdentity} can now be inserted into all identities corresponding to molecular coordinates such that,
\begin{equation}
   \hat h_{\text{A}i} = \1\otimes\underbrace{\left(\hat \theta_{\rm A} + \hat \theta_{\rm B}\right) \otimes ...\otimes 
   \left(\hat \theta_{\rm A} + \hat \theta_{\rm B}\right)}_{i-1 \text{ factors}} 
   \otimes\: \hat \theta_{\rm A} \otimes 
   \underbrace{\left(\hat \theta_{\rm A} + \hat \theta_{\rm B}\right) \otimes ... \otimes \left(\hat \theta_{\rm A} + \hat \theta_{\rm B}\right)}_{N-i \text{ factors}} \text{.}
\end{equation}
As each individual molecular Hamiltonian is the same, the labelling of particles does not matter.
% This means that the trace of the sum can firstly be expanded into a sum over the trace, followed by permutation of particles to arrange them in order, such that all $\hat \theta_{\rm A}$'s are next to each other. Then one can re-label the particles such that all traces are over the same basis and recombine the trace to combine all operators.
Therefore a rearrangement of the molecular projection operators is permitted such that,
\begin{equation}
    \hat h_{\rm A} = \frac{1}{N}\sum_{i=1}^N \1\otimes\underbrace{(\hat\theta_{\rm A}+\hat\theta_{\rm B})\otimes...\otimes(\hat\theta_{\rm A}+\hat\theta_{\rm B})}_{N-1 \text{ factors}}\otimes\:\hat\theta_{\rm A} \text{,}
\end{equation}
which is a sum over $N$ similar terms, allowing us to remove the sum and divide by $N$. Then, using the Binomial theorem,
\begin{equation}
    \hat h_{\rm A} =  \1\otimes
    \sum_{r=0}^{\!N\!-\!1\!}
    {\!N\!-\!1\!\choose r}
    \underbrace{\hat\theta_{\rm A}\otimes ... \otimes\hat\theta_{\rm A}}_{r+1 \text{ factors}}\otimes
    \underbrace{\hat\theta_{\rm B}\otimes ... \otimes\hat\theta_{\rm B}}_{N - (r+1) \text{ factors}} \text{,}
\end{equation}
which after rearranging can be written more simply as
\begin{equation}
    \hat h_{\rm A} =  \1\otimes
    \sum_{r=0}^{N}
    {N\choose r}
    \frac{r}{N}
    \underbrace{\hat\theta_{\rm A}\otimes ... \otimes\hat\theta_{\rm A}}_{r \text{ factors}}\otimes
    \underbrace{\hat\theta_{\rm B}\otimes ... \otimes\hat\theta_{\rm B}}_{N - r \text{ factors}}\text{.}
\end{equation}
The numerator of Eq.~\eqref{eqn:pANp} can therefore be expressed as an integral of the Boltzmann operator over a constrained space, just as in Eq.~\eqref{eqn:const.int} such that
\begin{equation}
\begin{split}
    \Tr{e^{-\beta\hat H}\hat h_{\rm A}} = \sum_{r=0}^{N}
    {N\choose r}
    \frac{r}{N}
    &
    \intinf dq_{\rm c} \prod_{j=1}^{r}\left(\int_{\!-\!\infty\!}^0 \!dq_j\right)\prod_{j=r+1}^{N}\left(\int_{0}^{\infty}\!dq_j\right)
 \\
 & \times\bra{q_{\rm c}\:q_{1}\:q_{2}\:...\:q_{N}}e^{-\beta\hat H}
 \ket{q_{\rm c}\:q_{1}\:q_{2}\:...\:q_{N}} \text{.}
\end{split}
\end{equation}
Now we can apply the same harmonic approximation as was made to yield Eq.~\eqref{eqn:harmaproxeq1}, but now over $N$ directions (corresponding to each integral over either A or B) such that
\begin{equation}
    \Tr{e^{-\beta \hat H}\hat h_{\rm A}} \approx \sum_{r=0}^{N}{N \choose r}
    \frac{r}{N}\Tr{e^{-\beta \hat H_{r} }}\text{,}
\label{eqn:TrA}
\end{equation}
where $\hat H_{r}$ is the Hamiltonian corresponding to a system of $N$ reaction coordinates, the first $r$ of which are on side A and the other $\!N\!-\!r\!$ of which are on side B, coupled to an optical cavity. 
\\
\indent 
Let us now focus on the potential of $\hat H_{\rm r}$, for which $N$ harmonic approximations have been made such that
\begin{equation}
\begin{split}
    \hat{V}_r = rE_{\rm A} + \frac{1}{2}\omega_{\rm A}^2\sum_{j=1}^r  \hat q_{j}^2
    +
    (\!N\!-\!r\!)&E_{\rm B} + \frac{1}{2}\omega_{\rm B}^2\sum_{j=r+1}^N \!\hat q_{j}^2
\\
+&
\frac{1}{2}\omega_{\rm c}^2\left(\hat q_{\rm c} + g_{\rm A}\sum_{j=1}^r \hat q_{j} + g_{\rm B}\sum_{j=r+1}^N \!\hat q_{j}\right)^2 \text{.}
\end{split}
\end{equation}
Via two unitary transforms, one can rotate into the `dark'/`bright' modes of coordinates confined in wells A and B. For those in A,
\begin{equation}
     \hat{q}_{\text{A}k} = \frac{1}{\sqrt{r}}\sum_{j=1}^r e^{\frac{2\pi \text{i}}{r}kj}\hat q_j \quad k=0,1,2\:...\:r\!-\!1 \text{,}
\end{equation}
and those in B,
\begin{equation}
     \hat{q}_{\text{B}k} = \frac{1}{\sqrt{N-r}}\sum_{j=1}^{N-r} e^{\frac{2\pi \text{i}}{N-r}lj}\hat q_{j+r} \quad k=0,1,2\:...\:N\!-\!r\!-\!1 \text{.}
\end{equation}
For both, $k=0$ corresponds to the totally symmetric linear combination of coordinates, which is commonly known as the `bright state' due to the fact this is the only linear combination that interacts with the optical cavity. As this is a unitary transformation, inner product is conserved such that the potential can be written in the new coordinates as
\begin{equation}
\begin{split}
    \hat{V}_r = rE_{\rm A} + \frac{1}{2}\omega_{\rm A}^2\sum_{k=0}^{r-1}  \hat q_{\text{A}k}^2
    +
    (\!N\!-\!r\!)&E_{\rm B} + \frac{1}{2}\omega_{\rm B}^2\sum_{k=0}^{\!N\!-\!r\!-\!1\!} \!\hat q_{\text{B}k}^2
\\
+&
\frac{1}{2}\omega_{\rm c}^2\left(\hat q_{\rm c} + g_A\sqrt{r}\:\hat q_{A0} + g_B\sqrt{N-r}\:\hat q_{B0}\right)^2 \text{.}
\end{split}
\end{equation}
This now means that $\hat H_{r}$ can be factored into
\begin{equation}
    \hat{H}_r = \sum_{k=1}^{r-1} \hat H_{\rm A}(\hat p_{\text{A}k},\hat q_{\text{A}k}) + \sum_{k=1}^{N\!-\!r\!-\!1} \hat H_{\rm B}(\hat p_{\text{B}k},\hat q_{\text{B}k})
    + \hat{H}^{\rm bright}_r
    % (\hat p_{\text{A}0},\hat p_{\text{B}0},\hat p_{\rm c},\hat q_{\text{A}0},\hat q_{\text{B}0},\hat q_{\rm c}) 
    \text{,}
\end{equation}
where $H_{\rm A}$ and $H_{\rm B}$ are the Hamiltonians of A and B for a single molecular system uncoupled from the optical cavity as defined in Eq.~\eqref{eqn:HarmHA}, and 
\begin{equation}
    \hat{H}^{\rm bright}_r = \frac{\hat p^2_{\text{A}0}}{2} + \frac{\hat p^2_{\text{B}0}}{2} + \frac{\hat p^2_{c}}{2} + \hat{V}^{\rm bright}_r \text{,}
\end{equation}
in which
\begin{equation}
\begin{split}
    \hat{V}^{\rm bright}_r =  E_{\rm A} + \frac{1}{2}\omega_{\rm A}^2 \hat q_{\text{A}0}^2 &+ E_{\rm B} + \frac{1}{2}\omega_{\rm B}^2 \hat q_{\text{B}0}^2 +
\\
 \frac{1}{2}\omega_{\rm c}^2&\left(\hat q_{\rm c} + g_\text{A}\sqrt{r}\: \hat q_{\text{A}0} + g_\text{B}\sqrt{N\!-\!r}\:\hat q_{\text{B}0}\right)^2 \text{.}
 \label{eqn:VrNmol}
\end{split}
\end{equation}
Inserting Eq.~\eqref{eqn:VrNmol} back into the Eq.~\eqref{eqn:TrA}, tracing over all uncoupled dark states - yielding the molecular partition functions $Q_{\rm A} = \Tr{e^{-\beta\hat H_{\rm A}}}$ and $Q_{\rm B} = \Tr{e^{-\beta\hat H_{\rm B}}}$ - we have 
\begin{equation}
        \Tr{e^{-\beta \hat H}\hat h_{\rm A} } \approx  \sum_{r=0}^{N}{N \choose r}
    \frac{r}{N}Q_{\rm A}^{r\!-\!1\!}Q_{\rm B}^{N\!-\!r\!-\!1\!}\Tr{e^{-\beta \hat{H}^{\rm bright}_r }} \text{.}
\end{equation}
The concentration of B can be expressed in an analogous form such that the equilibrium constant is
\begin{equation}
    K = \frac{\sum\limits_{r=0}^{N}{N \choose r}r\:
    Q_{\rm B}^{r\!-\!1}Q_{\rm A}^{N\!-\!r\!-\!1}\Tr{e^{-\beta  \hat{H}^{\rm bright}_{N-r}}}}{\sum\limits_{r=0}^{N}{N \choose r}r\:
    Q_{\rm A}^{r\!-\!1}Q_{\rm B}^{N\!-\!r\!-\!1}\Tr{e^{-\beta  \hat{H}^{\rm bright}_{r}}}} \text{.}
\end{equation}
Now let us introduce $r^\prime = N-r$ into the upper sum,
\begin{equation}
    K = \frac
    {\sum\limits_{r^\prime=0}^{N}{N \choose N-r^\prime}(N-r^\prime)\:
    Q_{\rm B}^{N\!-\!r^\prime\!-\!1}Q_{\rm A}^{r^\prime\!-\!1}\Tr{e^{-\beta  \hat{H}^{\rm bright}_{r^\prime}}}}
    {\sum\limits_{r=0}^{N}{N \choose r}r\:
    Q_{\rm A}^{r\!-\!1}Q_{\rm B}^{N\!-\!r\!-\!1}\Tr{e^{-\beta  \hat{H}^{\rm bright}_r}}}\text{.}
\end{equation}
Removing the prime, and cancelling the factors of $Q_{\text{A}}$ and $Q_{\text{B}}$,
\begin{equation}
    K = \frac{\sum\limits_{r=0}^{N}{N \choose r}(N-r)\left(\frac{Q_{\rm A}}{Q_{\rm B}}\right)^r\Tr{e^{-\beta  \hat{H}^{\rm bright}_r}}}
    {\sum\limits_{r=0}^{N}{N \choose r}r\left(\frac{Q_{\rm A}}{Q_{\rm B}}\right)^r\Tr{e^{-\beta  \hat{H}^{\rm bright}_{r}}}} \text{. }
\label{eqn:kNwTrHb}
\end{equation}
In analogy with the previous section, we diagonalise the Hessian of the `bright-mode' Hamiltonian
\begin{equation}
    \mt{V^{\prime\prime}_{r}} = \left(\begin{matrix}
    \omega_{\rm A}^2 + r\omega_{\rm c}^2g_{\rm A}^2 & \sqrt{r(N-r)}g_{\rm A}g_{\rm B}\omega_{\rm c}^2 &
    \sqrt{r}g_{\rm A}\omega_{\rm c}^2\\[0.5ex] 
    \sqrt{r(N-r)}g_{\rm A}g_{\rm B}\omega_{\rm c}^2   & \omega_{\rm B}^2 + (N-r)\omega_{\rm c}^2g_{\rm B}^2&
    \sqrt{N-r}g_{\rm B}\omega_{\rm c}^2\\[0.5ex] 
    \sqrt{r}g_{\rm A}\omega_{\rm c}^2   & \sqrt{N-r}g_{\rm B}\omega_{\rm c}^2&
    \omega_{\rm c}^2
    \end{matrix}\right) \text{,}
\label{eqn:HessianBright}
\end{equation}
in the basis
\begin{equation}
    \ket{q_{\rm A0}\: q_{\rm B0}\:q_{\rm c}} = 
    \ket{q_{\rm A0}}\otimes\ket{q_{\rm B0}}\otimes\ket{q_{\rm c}} \text{,}
\end{equation}
to yield eigenvalues $\{\lambda_{r1}^2,\lambda_{r2}^2,\lambda_{r3}^2\}$. Using these, and the quantum harmonic oscillator partition functions just as in the previous section,
\begin{equation}
    \Tr{e^{-\beta  \hat{H}^{\rm bright}_r}} = \prod_{i=1,2,3}\left[\frac{1}{2\sinh(\beta\hbar\lambda_{ri}/2)}\right]e^{-\beta(E_{\rm A}+E_{\rm B})} \text{,}
\end{equation}
which can be inserted into Eq.~\eqref{eqn:kNwTrHb} so that 
\begin{equation}
        K = \frac{\sum\limits_{r=0}^{N}{N \choose r}(N-r)\left(\frac{Q_{\rm A}}{Q_{\rm B}}\right)^r\prod\limits_{i=1,2,3}\left[\frac{1}{\sinh(\beta\hbar\lambda_{ri}/2)}\right]}
    {\sum\limits_{r=0}^{N}{N \choose r}r\left(\frac{Q_{\rm A}}{Q_{\rm B}}\right)^r\prod\limits_{i=1,2,3}\left[\frac{1}{\sinh(\beta\hbar\lambda_{ri}/2)}\right]} \text{. }
\end{equation}
Finally, one can notice that the ratio of single molecule uncoupled partition functions is the equilibrium constant of the reaction outside of a cavity, $K_0$ giving our final expression
\begin{equation}
            K = \frac{\sum\limits_{r=0}^{N}{N \choose r}\:(N-r)\:K_0^{-r}B(r)}
    {\sum\limits_{r=0}^{N}{N \choose r}\:r\:K_0^{-r}B(r)} \text{, }
\label{eqn:KnpFinalExp}
\end{equation}
where
\begin{equation}
    B(r)= \prod\limits_{i=1,2,3}\left[\frac{1}{\sinh(\beta\hbar\lambda_{ri}/2)}\right] \text{.}
\end{equation}
In Appendix~\ref{app:K0proof} it is shown that Eq.~\eqref{eqn:KnpFinalExp} reduces to $K_0$ when uncoupled from an optical cavity. The classical $\beta\to0$ limit of $K$,  
\begin{equation}
        K^{\rm cl} = \frac{\sum\limits_{r=0}^{N}{N \choose r}(N-r)K_0^{-r} \left[\frac{1}{\det \mt{V_r^{\prime\prime}}}\right]^{\frac{1}{2}}}
    {\sum\limits_{r=0}^{N}{N \choose r}rK_0^{-r} \left[\frac{1}{\det \mt{V_r^{\prime\prime}}}\right]^{\frac{1}{2}}} \text{, }
\end{equation}
which, due to the fact that
\begin{equation}
    \det \mt{V_r^{\prime\prime}} = \omega_{\rm A}^2\:\omega_{\rm B}^2\:\omega_{\rm c}^2 \text{,}
\end{equation}
can be simplified to 
\begin{equation}
    K^{\rm cl} = K_0 \text{.}
\end{equation}
Therefore - as expected from the single reactant case - the optical cavity will have no effect on the equilibrium constant  of a system of $N$ reactions in the classical limit. This observation  is corroborated by the findings of Li \textit{et al.\ }in Ref.~\cite{TaoLieqm2020}, who gave a proof that in the classical limit the cavity will not effect any static equilibrium properties of a system within. For the quantum case though there may be some effect, according to Eq.~\eqref{eqn:KnpFinalExp}. This will now be numerically tested.
% \section{\label{5:theory1}$N$-Particle equilibria}
% \input{text/5.Equilibrium/anotherproof}
% \section{\label{5:theory2}Path Integrals: Integrating out the cavity}
% \input{text/5.Equilibrium/OLDproof}
\section{\label{5:testingEqm}Simulation}
\begin{figure}[!h]
\centering
\includegraphics[trim={0.25cm 0.25cm 0.cm 0.25cm}]
{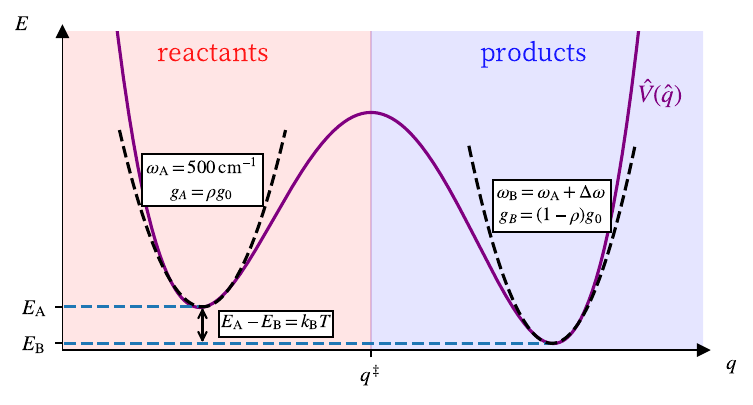}
\caption[Equilibrium system setup]{Parameterised equilibrium system setup, showing the harmonic approximations (dashed black lines) to the reaction coordinate potential $\hat V(\hat q)$ (purple line) at the reactant and product minima.}
\label{fig:eqmsetup}
\end{figure}

One could consider the cavity effects on the equilibrium constant of the symmetric double--well from the previous chapters. However, due to the symmetry of the potential energy surface, $\omega_{\rm A} = \omega_{\rm B}$ and $g_{\rm A}=g_{\rm B}$, meaning that no effects could possibly be observed in this setup.
\\
\indent 
Instead let us consider an asymmetric model, in which the well frequencies differ by $\Delta\omega = \omega_{\rm B} -\omega_{\rm A}$ and the coupling strength mismatch between reactants and products is parameterised by $0\leq\rho\leq1$, in which $\rho = 1$ corresponds to the products being uncoupled from the optical cavity and $\rho = 0$ the reactants being uncoupled. Fig.~\ref{fig:eqmsetup} shows this setup, along with the hard-coded values of $E_{\rm A} - E_{\rm A} = k_{\rm B}T$ and $\omega_{\rm A} = \SI{500}{cm^{-1}}$.
\FloatBarrier
\noindent
We define the coupling parameter as
\begin{equation}
    g_0 = 2\sqrt{\frac{2}{\omega_{\rm c}}}\eta_{\rm c} \text{,}
\end{equation}
which is chosen such that when $\rho = 0.5$, the Rabi splitting of reactants is unchanged from the previous sections for a given value of $\eta_{\rm c}$ (the same coupling constant was used by Lindoy \textit{et al.\ }in Ref.~\cite{Lindoy2022} and by us in Ch.~\ref{3:} and Ch.~\ref{4:}). When $\rho$ changes from $0$ to $1$, $\Omega_{\rm R}$ will change, but it remains on the order of $\SI{100}{cm^{-1}}$ in all cases we shall consider, ensuring that the system remains within the regime of vibrational strong coupling.
\\
\indent
Firstly let us consider a single molecule and probe the influence of $\Delta\omega$ and $\rho$ on the cavity effect on the equilibrium constant. The light-matter coupling constant is fixed at its maximum value from Sec.~\ref{3:PR} such that $\eta_{\rm c} = {5\times10^{-3}}{\rm a.u.}$ and the temperature is set to $\SI{300}{K}$.
\begin{figure}
\centering
\begin{subfigure}[c]{0.485\textwidth}
  \centering
\scalebox{0.95}{\includegraphics[width=\textwidth,trim={0cm 0.5cm 0.cm 0.8cm}]{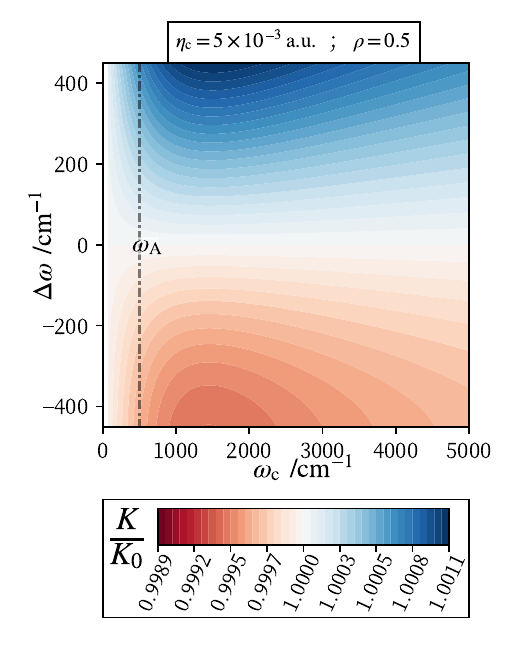}}
\end{subfigure}
\hfill
\vrule
\begin{subfigure}[c]{0.485\textwidth}
  \centering
  \scalebox{0.95}{\includegraphics[width=\textwidth,trim={0cm 0.5cm 0.cm 0.8cm}]{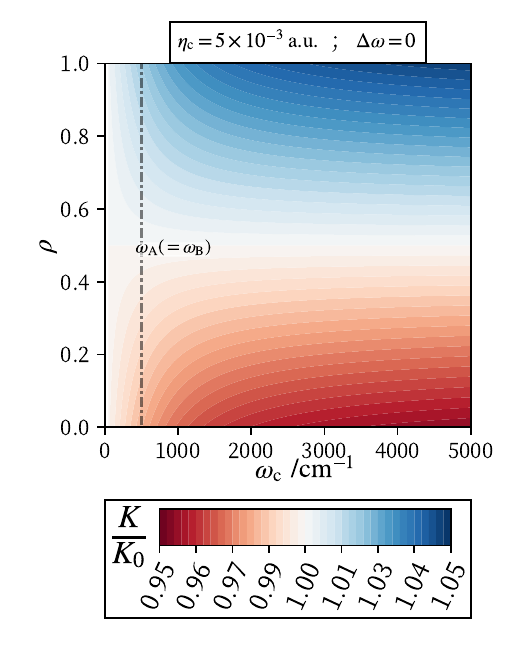}}
\end{subfigure}
\caption[Equilibrium properties parameter scan]{A scan over frequency miss-match (left panel) and polarity imbalance (right panel) to find conditions under which the optical cavity most affects the ratio of cavity-coupled to uncoupled equilibrium constants $K/K_0$ for a \textbf{single molecule reaction}. }
\label{fig:EqmPScan}
\end{figure}
In the left panel Fig.~\ref{fig:EqmPScan}, we explore the effect of changing $\Delta\omega$ whilst fixing $g_{\rm A} = g_{\rm B}$. For the right panel, we instead vary $\rho$ such that $g_{\rm A}\neq g_{\rm B}$, for $\omega_{\rm A}=\omega_{\rm B}$.
Both show weak effects (quantified by $K/K_0$) over a broad range of $\omega_{\rm c}$, with both enhancement and reduction of equilibrium constant observed. Now by selecting parameters that maximise the single molecule effect, let us see whether the observed effects survive in the collective regime. As always, we will keep $\eta_{c}\sqrt{N}$ constant, such that the Rabi splitting is constant as $N$ increases.
% \\
% \indent
% Both equilibrium constant enhancement and suppression are observed, the trend of which is reversed when the well height imbalance, $E_{\rm A} - E_{\rm B}$, is reversed. For these plots, we have fixed $\rho = 0.5$ for left panel (as this was found to give least effect for any given $\Delta\omega$) and $\Delta\omega = 0$ for right panel (analogous for any given $\rho$). 
\begin{figure}[!h]
\centering
\includegraphics[trim={0.25cm 0.75cm 0.cm 0.8cm}]
{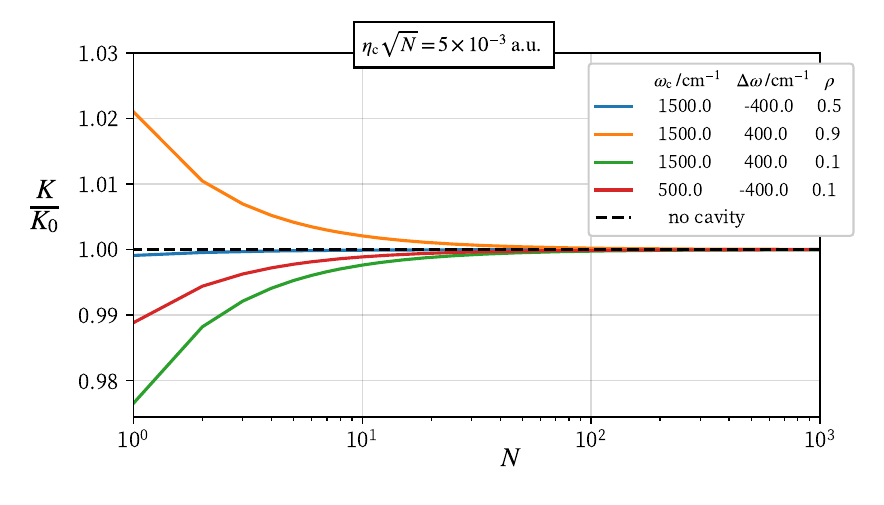}
\caption[Equilibrium effects in the collective regime]{Cavity effects on the equilibrium constant `scale out' as $N$ is increased. The lines are chosen to illustrate that this is seen throughout the (very large) parameter space.}
\label{fig:eqmColeff}
\end{figure}
% \FloatBarrier
\\
\indent 
Fig.~\ref{fig:eqmColeff} shows the same `scaling-out' behaviour as previously observed in Fig.~\ref{fig:ResNP}. Note that in preliminary tests very many possible values of $\Delta\omega$, $\omega_{\rm c}$ and $\rho$ were tested - all with the same  results: any effect on single molecule equilibrium constants observed is negligible in the collective regime. 
\\
\indent 
Though the derivation of Sec.~\ref{5:theoryNP} applied for a single molecular degree of freedom, the multidimensional generalisation is simple. It only requires replacing $\omega_{\rm  A}^2$ and $\omega_{\rm B}^2$ with the Hessians of the multidimensional system, and a vectorising of the coupling constants $g_{\rm A}$ and $g_{\rm B}$. By including solvent bath degrees of freedom within each system, friction effects on the reaction coordinate could be included. In addition, a cavity loss bath could be added due to there being no direct interaction between this bath and the reaction coordinates, allowing the dark states to be `integrated out', just as we have done here. When at equilibrium though, these baths will be in equilibrium with the systems they are attached to, slightly altering the effective frequency distribution of the systems but otherwise having no significant effect. 
\\
\indent 
For these reasons - even if friction were to be added - the single-molecule effects would scale out in the same way as shown in Fig.~\ref{fig:eqmColeff}, hence we will stop our analysis of  equilibrium effects here. This decision is further justified by considering the results of Fig.~\ref{fig:res_SD_1P}, in which it was observed that the inclusion of friction and loss did not significantly effect the transition state theory rate of reaction relative to Yang and Cao's results \cite{Yang2021}. 
\\
\indent 
This concludes our examination of polariton chemistry. Next chapter we will summarise our findings, discussing the possible insights this thesis has given towards the understanding of chemistry under vibrational strong coupling.

% \section{\label{5:eqmresults}Results}
% \input{text/5.Equilibrium/EqmResults}

\chapter{\label{Conclusion}Conclusion} 
\minitoc
\indent 
In this thesis we have used various theories/methods to study the effect of vibrational strong coupling on chemical reaction dynamics. Let us firstly summarise the results of the preceding chapters:
\begin{itemize}
\item[-] Ch.~\ref{3:} addressed the dynamics of a single reaction coupled to the optical cavity. The origin of the rate modification observed here was found to be a dynamical recrossing effect, present only in the low-friction regime. For this reason, transition state theory was concluded to be insufficient to explain the observed effects (see Sec.~\ref{3:Conclusion}).
\item[-]  Ch.~\ref{4:} then demonstrated that none of the effects found for the single-reaction study would survive in the limit of a macroscopic number of $N$ reactions collectively coupled to the cavity.
\item[-]  Ch.~\ref{5:} turned the focus towards cavity effects on equilibria. In contrast with the previous findings, these were found to be solely quantum-mechanical. The effects observed did again however succumb to the same problem at large $N$, indicating that no significant change in rate or equilibrium constant would be observed for an experimentally realistic value of $N$.
\end{itemize}
A large, and ever growing number of theoretical studies \cite{Lindoy2022,anderson2023,philbin2022} have now been published discussing single reactions coupled to optical cavities. The analysis in this thesis (in agreement with Ref.~\cite{JoelReview}) highlights a common problem with many of these studies: under the Pauli--Fierz Hamiltonian, the optical cavity couples to the total dipole of the system, summed over each individual reactant. 

This means that when $N$ \textbf{identical} quantum systems are collectively coupled to an optical cavity, a single linear combination of their individual dipoles couples to the cavity mode (forming the upper and lower polaritons) leaving $N-1$ dark, \textbf{completely unaffected} combinations unchanged. For this reason no effects observed for a single molecule remain in the collective regime, in agreement with our findings from Chs.~\ref{4:} and~\ref{5:}: single molecule effects may be observed, but as $N$ increases the dark states dominate the system's characteristics leading to the scaling-out observed. Note that if no rates are affected, then by detailed balance, equilibrium constants also cannot be effected.
\\
\indent
This analysis raises the question of whether there does exist some mechanism by which dark states could be modified by vibrational strong coupling. Botzung \textit{et al.\ }in Ref.~\cite{Semilocalisation} used Markovian master equation treatment to study an ensemble of \textbf{disordered} two-level systems (each of slightly different transition frequency) collectively coupled to an optical cavity. They found that whilst still forming a lower and upper polariton, this disorder also led to hybridisation between dark states of similar energies, allowing coherent energy transportation between them, even in the thermodynamic (large $N$) limit. This may imply that the assumption of identical reactive systems was a shortcoming in our investigations, and that for collective effects to be observed both a quantum-mechanical treatment of systems (for proper inclusion of coherences) and disorder amongst these systems (to promote dark state mixing) is required. In addition, it may also be possible - in analogy with disorder-driven mixing - that mixing could occur between vibrational transitions of similar energy for anharmonic potentials, if treated quantum-mechanically.
\\
\indent 
Another possible shortcoming of the present study could be the single mode approximation. The results of Figs.~\ref{fig:res_ED_1P},~\ref{fig:res_SD_1P} and~\ref{fig:EqmPScan} do not show any well defined polariton resonance, but instead effects over a large spectrum of cavity frequencies $\omega_{\rm c}$. This could imply that there are interactions between non-resonant cavity modes and reactive systems. Could it be that rate/equilibrium constant modification can only be observed as the combined interaction of all of these modes with the system?
\\
\\
\indent
Since it has proven so difficult to explain the rate effect, it is worth carefully analysing the existing experimental evidence for this phenomenon. There have so far been at least two failed reproductions of the motivational experimental evidence.
In Ref.~\cite{repro1}, Imperatore \textit{et al.\ }found no cavity effect on the rate of cyanate ion hydrolysis despite observing the same Rabi splitting as Hiura \textit{et al.\ }\cite{Hiura2018}, who previously observed a 100-fold rate increase during their experiments. Also, Wiesehan and Xiong tried, and failed~\cite{repro2}, to reproduce the 10-fold rate increase observed for ester hydrolysis of para-nitrophenyl acetate by Lather \textit{et al.\ }\cite{h2oSolvCoup}, again despite observing the same Rabi splitting.
\\
\indent
In addition to the lack of reproducibility, issues regarding measurement with optical methods have been raised.
Di Virgilio \textit{et al.\ }explored the cavity influence on the electro-optic response of a semiconducting perovskite coupled to resonant optical cavity ~\cite{BonnPaper}. Despite finding no effects on the material properties, they did find significant changes in the spectroscopic response of perovskite coupled to the cavity. Spectroscopic techniques such as  monitoring the infrared transmission spectrum of higher-order cavity modes \cite{Hirai2020}, or detecting the reactant/product concentration via UV/Visible spectroscopy \cite{EqmCavity}, are commonly used to measure the extent of reactions occurring within optical cavities. The findings of Di Virgilio \textit{et al.\ } open the possibility that the results could be purely an optical effect, and not rate modification as reported. In addition to this, as these are `pump-probe' type experiments, could it be that the measurement of reaction extent by these methods is resulting in an inadvertent pumping of the optical cavity with extra energy?
\\
\indent
Whilst this thesis may have only shown that the Hamiltonian used is insufficient to explain observed polariton effects on reaction rates, the ubiquitous failure of other theoretical explanations could instead be seen as further evidence undermining the validity of the experimental findings. Until the speculations regarding the reproducibility of the experimental results are resolved, there is therefore no answer as to whether chemistry under vibrational strong coupling is a triumph of modern science or simply a synthetic chemist's dream.

{\footnotesize
\setlength{\baselineskip}{0pt}
\renewcommand*\MakeUppercase[1]{#1}
\bibliography{refs}{}}

% \include{text/Tools/tools}
% \include{text/ROC/ROC}

% {\footnotesize\setlength{\baselineskip}{0pt}\renewcommand*\MakeUppercase{1}{#1}\bibliography{references}{}}
%% APPENDICES %% 
% Starts lettered appendices, adds a heading in table of contents, and adds a
%    page that just says "Appendices" to signal the end of your main text.
\startappendices
% Add or remove any appendices you'd like here:
% \include{text/appendix-1}
% \include{text/A1-StaticEqmProperties}
% \include{text/A2-Kramer'sTheory}
\chapter{\label{app:IneEOM}The Cayley Propagator}
When numerically integrating harmonic equations of motion, it is usually a rule of thumb that the time-step $\delta t$ must be sufficiently small such that,
\begin{equation}
    \delta t \leq \frac{0.1\pi}{\omega_{\rm max}} \text{,}
\end{equation}
in which $\omega_{\rm max}$ is the maximum frequency of the system. When integrating the equations of motion of a ring polymer, this results in an annoying feature for simulations with large numbers of beads, as high frequency normal mode components require very small time steps for integration to remain stable. This appendix describes a fix to this problem that is very effective for the special (system + bath) case considered in this thesis.
\section{\label{app:Cayley}Cayley Propagator}
In this description, a single molecular degree of freedom (DOF), connected to its own Caldiera-Leggett bath will be considered. Multidimensional generalisation is as straightforward as adding another index to all variables. Also, for simplicity we use mass-scaled coordinates and momenta. The structuring of the variables is as follows.
\begin{enumerate}
    \item[-] There are $n_{\rm f}$ physical degrees of freedom. The first index corresponds to the molecular DOF, the remaining indices from $2$ to $n_{\rm f}$ correspond to those of a discretised bath with $n_{\rm f}-1$ modes (see Appendix~\ref{app:3-}).
    \item[-] $\omega_{i}$ and $c_{i}$ hold the discretised bath frequencies/coupling strengths, each in an array of size $n_{\rm f}$. The index from $2$ to $n_{\rm f}$ hold the information for each the $n_{\rm f}-1$ bath modes. 
    The first index is fixed such that $c_1 = \omega_1^2$, such that the summations in Eq.~\eqref{eqn:H0-rpbath} may be over all molecular DOFs, simplifying notation.
    \item[-] The phase-space variables $p_{ij}$ and $q_{ij}$ have two indices. The $i$ index runs from $1$ to $n_{\rm f}$ and corresponds to the physical DOFs. The $j$ index is the ring polymer bead index. It runs form $1$ to $n_{\rm b}$, the number of ring polymer beads. 
    \item[-] $\omega_{n} = \frac{1}{\beta_n\hbar}$ is the ring polymer spring frequency.
\end{enumerate}
Using this structure, let us setup the ring polymer Hamiltonian of a generic 1-dimensional system coupled to a bath of harmonic oscillators such that,
\begin{equation}
    H(\{p_{ij}\},\{q_{ij}\}) = H_0(\{p_{ij}\},\{q_{ij}\}) + V(\{q_{1j}\}) \text{.}
\end{equation}
Here 
\begin{equation}
    V(\{q_{1j}\}) = \sum_{j=1}^{n_{\rm b}} V(q_{1j}) \text{,}
\end{equation}
the potential energy experienced by the molecular system (which is not necessarily harmonic, and hence separated from $H_0$) and
\begin{equation}
\begin{split}
\\[0.1cm]
    H_0(\{p_{ij}\},\{q_{ij}\}) = \sum_{i=1}^{n_{\rm f}}\sum_{j=1}^{n_{\rm b}}\left(\frac{p_{ij}^2}{2}+ \smash{\overbrace{\frac{1}{2}\omega_{n}^2(q_{ij}-q_{ij\!+\!1})^2}^{\text{Ring polymer\:springs}\:V_{\rm rp}}}
    +
 \smash{\underbrace{\frac{1}{2}\omega_{i}^2\left(q_{ij}-\frac{c_i}{\omega_i^2}q_{1j}\right)^2}_{\text{System-bath\: interactions }V_{\rm sb}}}\:
    \right) \text{.}
    \label{eqn:H0-rpbath}
\\[0.75cm]
\end{split}
\end{equation}
As the ring polymer potential is the same for all $i$ and the system bath  potential for all $j$, it is useful to define the ring polymer Hessian 
\begin{equation}
    (\mt{V_{\rm rp}^{\prime\prime})}_{jj'} = \frac{\partial^2V_{\rm rp}}{\partial q_{ij}\partial q_{ij'}} \text{,}
\end{equation}
and the system-bath Hessian
\begin{equation}
    (\mt{V_{\rm sb}^{\prime\prime}})_{ii'} = \frac{\partial^2V_{\rm sb}}{\partial q_{ij}\partial q_{i'j}} \text{,}
\end{equation}
as they allow us to re-write Eq.~\eqref{eqn:H0-rpbath} in terms of $n_{\rm f}$ by $n_{\rm b}$ matrices $\mt{p}$ and $\mt{q}$ as
\begin{equation}
\begin{split}
        H_0(\mt{p},\mt{q}) = \frac{1}{2}\left(\Tr{\mt{p^\top}\mt{p}}
+
\Tr{\mt{q}\mt{V_{\rm rp}^{\prime\prime}}\mt{q^\top}}
         +       \Tr{\mt{q^\top}\mt{V_{\rm sb}^{\prime\prime}}\mt{q}} \right)\text{.}
         \label{eqn:H0rewritten}
\end{split}
\end{equation}
Let us define the unitary transformations that diagonalise $\mt{V_{\rm rp }^{\prime\prime}}$ and $\mt{V_{\rm sb}^{\prime\prime}}$ such that,
\begin{equation}
    \mt{V_{\rm rp}^{\prime\prime}}
    = \mt{U_{\rm rp}}\:\mt{\omega^2_{\rm k}}\mt{U_{\rm rp}^\top} \text{,}
\end{equation}
 in which $\mt{\omega^2_{\rm k}}$ is a diagonal matrix of eigenvalues $\omega_k^2$ and
\begin{equation}
    \mt{V_{\rm sb}^{\prime\prime}}
    = \mt{U_{\rm sb}}\:\mt{\omega^2_{\rm l}}\mt{U_{\rm sb}^\top} \text{,}
\end{equation}
in which $\mt{\omega^2_{\rm l}}$ is a diagonal matrix of eigenvalues $\omega_l^2$. Therefore by two unitary transforms,  Eq.~\eqref{eqn:H0-rpbath} can be re-written as
\begin{equation}
\begin{split}
    H_0(\{\tilde p_{lk}\},\{\tilde q_{lk}\}) = \sum_{l=1}^{n_{\rm f}}\sum_{k=1}^{n_{\rm b}}\left(\frac{\tilde p_{lk}^2}{2}+ \frac{1}{2}\smash[b]{\underbrace{\left(\omega_l^2+\omega_k^2\right)}_{\omega^2_{lk}}} \tilde q_{lk}^2
    \right) \text{.}
\end{split}
\end{equation}
where
\begin{equation}
    \tilde q_{lk} = \sum_{i=1}^{n_{\rm f}}\sum_{j=1}^{n_{\rm b}}(\mt{U_{\rm sb}})_{il}\: q_{ij}\:(\mt{U_{\rm rp}})_{jk} \text{,}
\end{equation}
or (equivalently)
\begin{equation}
    \mt{\tilde q} = \mt{U_{\rm sb}^\top}\: \mt{q} \:\mt{U_{\rm rp}} \text{.}
\end{equation}
The momentum is transformed similarly. For the element $lk$ in this new basis, the equations of motion generated by the Hamiltonian can be written in matrix form as
\begin{equation}
    \frac{\rm d}{\text{d}t}
    \underbrace{\begin{pmatrix}\tilde p_{lk}\\ \tilde q_{lk} \end{pmatrix}}_{\vc{\tilde x_{lk}}(t)}
=
\underbrace{\begin{pmatrix}
0 & -\omega_{lk}^2 \\
1 & 0
\end{pmatrix}}_{\mt{A_{lk}}}
\begin{pmatrix}\tilde p_{lk}\\ \tilde q_{lk} \end{pmatrix}
\end{equation}
which can be numerically integrated using
\begin{equation}
    \vc{\tilde x_{lk}}(t\!+\!\delta t) = e^{\mt{A_{lk}}\delta t} \:   \vc{\tilde x_{lk}}(t)
\end{equation}
The Cayley propagator uses an approximation to the exponential, which is more stable for larger time-steps when used within the split-operator scheme in Eq.~\eqref{eqn:spltopscheme} below \cite{cayley}. The approximation itself is
\begin{equation}
 e^{\mt{A_{lk}}\delta t} \approx \left(\mathds{1}- \mt{A_{lk}}\frac{\delta t}{2}\right)^{-1}
 \left(\mathds{1} + \mt{A_{lk}}\frac{\delta t}{2}\right)\text{.}
\end{equation}
We can now insert the definition of $\mt{A_{lk}}$ to obtain an analytical expression for the propagator,
\begin{equation}
 \begin{split}
      e^{\mt{A_{lk}}\delta t} &\approx 
\begin{pmatrix}
1 & \omega_{lk}^2\frac{\delta t}{2} \\[0.1cm]
-\frac{\delta t}{2} & 1
\end{pmatrix}^{-1}
\begin{pmatrix}
1 & -\omega_{lk}^2\frac{\delta t}{2} \\[0.1cm]
\frac{\delta t}{2} & 1
\end{pmatrix}
\\[0.5cm]
&= \frac{1}{1+c_{lk}} \begin{pmatrix}
1-c_{lk} & -\omega_{lk}^2\delta t \\[0.1cm]
\delta t & 1-c_{lk}
\end{pmatrix}\text{,}
 \end{split}   
\end{equation}
where we have defined $c_{lk} = \frac{\omega_{lk}^2\delta t^2}{4}$. With this propagator we can apply the Liouvillian $\mathcal{L}_0$ corresponding to $H_0$, ensuring that the integration is stable. To integrate the equations of motion generated by $H$ though, the external potential must also be taken into account. We do this by symmetrically splitting the integration such that,
\begin{equation}
    e^{\mathcal{L}\delta t} \approx e^{\mathcal{L}_V\frac{\delta t}{2}}\:
    \underbrace{e^{\mathcal{L}_0 \delta t}}_{\rm Cayley}\:
    e^{\mathcal{L}_V\frac{\delta t}{2}} \text{.}
    \label{eqn:spltopscheme}
\end{equation}
With the Cayley propagator, integration is done in the normal mode coordinates (tildes) of the phase-space variables. For $\mathcal{L}_V$ though, the potential must be evaluated using the physical coordinates (no tilde). As the external potential is only applied to $q_{1j}$, we only need to transform this component, calculate the change in the conjugate momenta $\delta p_{1j}$ over each half time step, then transform back to the normal mode representation.
\\
\\
To achieve $e^{\mathcal{L}_V\frac{\delta t}{2}}$, firstly obtain $\{q_{1j}\}$ via
\begin{equation}
    q_{1j} = \sum_{l=0}^{n_{\rm f}\!-\!1}\sum_{k=0}^{n_{\rm b}\!-\!1}(\mt{U_{\rm sb}})_{1l}\: \tilde q_{lk}\:(\mt{U_{\rm rp}})_{jk} \text{,}
\end{equation}
then using Hamilton's equations calculate
\begin{equation}
    \delta p_{1j} = - \frac{\delta t}{2}V'(q_{1j})\text{,}
\end{equation}
then transform back $\delta p$ and apply to the system
\begin{equation}
    \tilde p_{lk} \xleftarrow{} \tilde p_{lk} + \sum_{j=1}^{n_{\rm b}}(\mt{U_{\rm sb}})_{1l}\: \delta p_{1j}\:(\mt{U_{\rm rp}})_{jk}\text{.}
\end{equation}
This algorithm allows fast and stable integration of the equations of motion of any (adiabatic) ring polymerised system, in which the majority of the interactions are harmonic (as they are in the system-bath model).

\chapter{\label{app:3-}Bath Discretisation and GLEs}
The system--bath model is used to model the interactions between a system and its environment, for example a solvent. In this appendix, we describe how the baths were constructed. For each bath, the spectral density $J(\omega)$ (see Sec.~\ref{2:SBM}) was discretised into a sum over a finite number of modes,
\begin{equation}
    J_{n}(\omega) = \frac{\pi}{2} \sum_{i=1}^{n} 
    \frac{c_{i}^{2}}{\omega_{i}}\delta(\omega-\omega_i)
    \label{eqn:discret.}
\end{equation}
\section{\label{app:3-ohmic}Discrete Ohmic Baths}
The results of section \ref{2:KramTh} were reproduced from \cite{Craig2005}. In this paper an Ohmic bath was used with spectral density of the form
\begin{equation}
    J_{\rm Ohm.}(\omega) = \eta\omega e^{-\omega/\gamma}\text{,}
\end{equation}
where the cutoff frequency $\gamma$ and solvent strength $\eta$ is that of the bath in question.
The Ohmic bath was discretised into the form in Eq.~\eqref{eqn:discret.} through the same scheme as in \cite{Craig2005} which is
\begin{equation}
    \omega_{i} = -\gamma\ln\left(\frac{i - 1/2}{n}\right)
    \:\:\:\:\:\:\:\:\:\:\:
    c_{i} = \omega_{i}\left(\frac{2\eta\gamma}{n\pi}\right)^{\frac{1}{2}} \text{,}
\end{equation}
for $i = 1,2,...n$. For this model, $n = 9$ was required for graphical convergence (see abstract of Ref.~\cite{Craig2005}).
\section{\label{app:3-debye}Discrete Debye baths}
For the remainder of the thesis (Chapter~\ref{3:} onwards) Debye baths were used. For this,
% \begin{equation} OLD ONE
%     J_{\rm Deb.}(\omega) = \eta \frac{\omega\:\gamma}{\omega^2 +\gamma^2} \quad\text{,}
% \end{equation}
\begin{equation} %NEQ ONE
    J_{\rm Deb.}(\omega) = \eta \omega\:\frac{\gamma^2}{\omega^2 +\gamma^2} \text{,}
\end{equation}
where $\eta$ is the solvent strength and $\tau = 1/\gamma$  is timescale corresponding to the Debye spectral density.
Now the discretisation scheme to assign $c_{i}$ and $\omega_{i}$ for a given number of bath modes $n$ is
\begin{equation}
    \omega_{i} = \gamma\tan\left(\frac{(i-1/2)\pi}{2n}\right)
    \:\:\:\:\:\:\:\:\:\:\:
    c_{i} = \omega_{i}\left(\frac{\eta\gamma}{n}\right)^{\frac{1}{2}} \text{.}
    \:\:\:
\end{equation}
for $i = 1,2,...n$.
\section{\label{app:3-GLES}Generalised Langevin Equations (GLEs)}
For the classical simulations of Chapter~\ref{4:}, instead of including discretised baths, GLEs were employed in order to improve computational efficiency. For the Debye spectral density there is an efficient algorithm in terms of a Prony series which follows Ref.~\cite{Baczewski2013}. This is the same algorithm used by Lindoy \textit{et al. }in Ref.~\cite{Lindoy2022}, the implementation of which is discussed in their appendix.
\chapter{\label{app:4-}Model Parameters}
For Ch.~\ref{2:}, the model parameters used were the same as in Refs.~\cite{Topaler1994,Craig2005}. As this model is not the main focus of this thesis they will not be included here. 
\\
\\
For Chs.~\ref{3:} and ~\ref{4:} (for the same model used in Fig. S2 of the supplementary information of Ref.\cite{Lindoy2022} ) the following solvent cutoff parameters, double--well parameters and timestep were used,
\begin{center}
\begin{tabular}{||c|c|c|c|c||} 
 \hline
 $\gamma_{\rm L}$ & $\gamma_{\rm s}$ & $\omega_{\rm b}$ & $E_{\rm b}$& ${\rm dt}$ \\ [0.5ex] 
 \hline
 $\SI{1000}{cm^{-1}}$ & $\SI{200}{cm^{-1}}$ & $\SI{500}{cm^{-1}}$ & $\SI{2000}{cm^{-1}}$& $\SI{20}{au}$ \\ 
 \hline
\end{tabular}
\end{center}
For Chs.~\ref{3:} and~\ref{5:} the temperature was set to $\SI{300}{K}$, for Ch.~\ref{4:} however, the temperature was set to $\SI{500}{K}$. All other information for the parameters used in Ch.~\ref{5:} can be found in Sec.~\ref{5:testingEqm}.
\section{Convergence information: Flux-Side Correlation}
For all simulations, the number of bath modes, plateau time and number of trajectories was the same.
\begin{center}
\begin{tabular}{||c|c|c|c||} 
 \hline
 $n_{\rm L}$ & $n_{\rm S}$ & $t_{\rm p}$ & $n_{\rm traj.}$ \\ [0.5ex] 
 \hline
 $200$ & $200$ & $\SI{1200}{fs}$ & $2\times10^5$ \\ 
 \hline
\end{tabular}
\end{center}
These were sufficient to converge the results shown in the plots.
\section{Convergence information: Direct Simulation}
For the direct simulation, firstly the reactants were equilibriated for $\SI{2500}{fs}$, from an initial condition of all coordinates in the center of the reactant well, with their momenta sampled from the Maxwell--Boltzmann distribution. The system's coordinates and momenta was then recorded at $\SI{100}{fs}$ intervals, until enough samples had been taken. Then each snapshot was simulated for $\SI{5000}{fs}$ in stage $2$. The number of trajectories in the main part of Fig.~\ref{fig:ResNP} were
\begin{center}
\begin{tabular}{||c||c|c|c|c|c||} 
 \hline
 $n_{\rm sys.}$ & $1$ & $5$ & $10$ & $25$ & $50$ \\ [0.5ex] 
 \hline
 $n_{\rm traj.}$& $1\times10^5$ &$1\times10^5$ & $1\times10^5$ & $5\times10^4$ & $5\times10^4$ \\ 
 \hline
\end{tabular}
\end{center}
For the inset in Fig.~\ref{fig:ResNP}, the number of trajectories was calculated as $n_{\rm traj.} = 10^5/\sqrt{n_{\rm sys.}}$.
\chapter{\label{app:K0proof}Outside-cavity Equilibria}
If Eq.~\eqref{eqn:KnpFinalExp} is to be true it must reduce to the outside-cavity equilibrium constant $K_0$ when the coupling constants $g_{\rm A} = g_{\rm B} = 0$. Looking at the Hessian from Eq.~\eqref{eqn:HessianBright} by which $\{\lambda_{r1},\lambda_{r2},\lambda_{r3}\}$ are obtained, all terms containing $r$ are multiplied by either $g_{\rm A}$ or $g_{\rm B}$. Therefore when uncoupled, the eigenvalues will not have any $r$ dependence. This means the bright-state factor $B(r)$ is now a constant and as such may be cancelled such that the uncoupled equilibrium constant
\begin{equation}
            K_{\rm uncoup.} = \frac{\sum\limits_{r=0}^{N}{N \choose r}\:(N-r)\:K_0^{-r}}
    {\sum\limits_{r=0}^{N}{N \choose r}\:r\:K_0^{-r}}\text{. }
\label{eqn:KnPNOCAV}
\end{equation}
Looking at the denominator,
\begin{equation}
\begin{split}
    \sum\limits_{r=0}^{N}{N \choose r}\:r\:K_0^{-r} &= K_0^{-1}\sum\limits_{r=0}^{N}{N \choose r}\:r\:K_0^{-r+1}
    \\
    &= K_0^{-1}\sum\limits_{r=1}^{N}\frac{N!}{r!(N-r)!}\:r\:K_0^{-r+1}
    \\
    &= K_0^{-1}\sum\limits_{r=1}^{N}\frac{N!}{(r-1)!(N-(r-1))!}\:(N-(r-1))\:K_0^{-(r-1)} \text{.}
\end{split}
\end{equation}
Now let us introduce $r^\prime = r-1$, such that
\begin{equation}
\begin{split}
    \sum\limits_{r=0}^{N}{N \choose r}\:r\:K_0^{-r} &= K_0^{-1}\sum\limits_{r^\prime=0}^{N-1}\frac{N!}{(r^\prime)!(N-r^\prime)!}\:(N-r^\prime)\:K_0^{-r^\prime}
    \\
    &= K_0^{-1}\sum\limits_{r^\prime=0}^{N}\frac{N!}{(r^\prime)!(N-r^\prime)!}\:(N-r^\prime)\:K_0^{-r^\prime}
    \\
    &= K_0^{-1}\sum\limits_{r=0}^{N}{N\choose r}\:(N-r)\:K_0^{-r} \text{,}
\end{split}
\end{equation}
in which the prime was removed in the last step (as labelling was arbitrary). Inserting this back into Eq.~\eqref{eqn:KnPNOCAV},
\begin{equation}
\begin{split}
    \frac{\sum\limits_{r=0}^{N}{N \choose r}\:(N-r)\:K_0^{-r}}
    {\sum\limits_{r=0}^{N}{N \choose r}\:r\:K_0^{-r}} &=      \frac{\sum\limits_{r=0}^{N}{N \choose r}\:(N-r)\:K_0^{-r}}
    {K_0^{-1}\sum\limits_{r=0}^{N}{N\choose r}\:(N-r)\:K_0^{-r}}
    \\
    &=K_0 
    \\
    &\therefore K_{\rm uncoup.} = K_0 \text{.}
\end{split}
\end{equation}
%%%%% REFERENCES

% JEM: Quote for the top of references (just like a chapter quote if you're using them).  Comment to skip.
% \begin{savequote}[8cm]
% The first kind of intellectual and artistic personality belongs to the hedgehogs, the second to the foxes \dots
%   \qauthor{--- Sir Isaiah Berlin \cite{berlin_hedgehog_2013}}
% \end{savequote}

\end{document}